\documentclass[pdflatex,sn-mathphys-num]{sn-jnl}% Math and Physical Sciences Numbered Reference Style
\usepackage{graphicx}%
\usepackage{multirow}%
\usepackage{amsmath,amssymb,amsfonts}%
\usepackage{amsthm}%
\usepackage{mathrsfs}%
\usepackage[title]{appendix}%
\usepackage{xcolor}%
\usepackage{textcomp}%
\usepackage{manyfoot}%
\usepackage{booktabs}%
\usepackage{algorithm}%
\usepackage{algorithmicx}%
\usepackage{algpseudocode}%
\usepackage{listings}%
\usepackage{colortbl}
\usepackage{longtable}
\usepackage{setspace}
\theoremstyle{thmstyleone}%
\theoremstyle{thmstyletwo}%

\theoremstyle{thmstylethree}%

\begin{document}

\definecolor{lightgray}{gray}{0.9}

\title[Article Title]{Towards a better approach to the Vehicle Routing Problem}

\author[]{\fnm{Abdoune} \sur{Souad} and \fnm{Boulif} \sur{Menouar}} \email{s.abdoune@univ-boumerdes.dz, {m.boulif@univ-boumerdes.dz, boumen7@gmail.com}}

\affil[]{ LIMOSE laboratory, \orgdiv{Department of Computer Science, Faculty of science}, \orgname{University Mhamed Bouguerra of Boumerdes}, \city{Boumerdes}, \country{Algeria}}


  \abstract{ {\color{black}The Vehicle Routing Problem (VRP) is a fundamental challenge in logistics management research, given its substantial influence on transportation efficiency, cost minimization, and service quality. As a combinatorial optimization problem, VRP plays a crucial role in a wide range of real world applications, particularly in transportation, logistics, and delivery systems, due to its diverse formulations and numerous extensions. Over the years, researchers have introduced various VRP variants to address specific operational constraints, emerging industry requirements and optimize specific objectives, making it one of the most extensively studied problems in operations research. This article provides a comprehensive overview of VRP by exploring its theoretical foundations, discussing the limitations of its classical model, and introducing its key extensions. By systematically reviewing the diverse constraints, objectives, and variants examined in recent literature, this study aims to contribute to a deeper understanding of VRP while highlighting its ongoing evolution and relevance in modern optimization and decision making processes.}}

\keywords{Optimization, Modeling, Vehicle Routing Problem, Route planning}


\pacs[MSC Classification]{90c59}

\maketitle

\section{Introduction}\label{sec1}

The Vehicle Routing Problem was presented the first time by Dantzig and Ramser \cite{Braekers2016} to model how truck groups can serve the oil demands to a set of gas stations from a terminal, while minimizing the total traveled distance. It is noteworthy that the VRP stands as a generalization of the well-known Traveling Salesman Problem (TSP) \cite{Dantzig1959, Boulif2024}. Indeed, while the TSP focuses on finding the shortest possible route for a single salesman to visit a set of cities and return to the starting point, the VRP extends this definition by incorporating multiple vehicles and accounting for limited vehicle capacities \cite{Laporte2007}.

VRP is a combinatorial optimization problem that, in its basic form, aims to define the optimal routes for homogeneous fleet of vehicles to serve predefined customers at different locations from a single depot. Each customer must be visited once by only one vehicle \cite{Bansal2018}. Actually, VRP has become one of the most interesting topics to combinatorial optimization researchers due to its vital role in solving real-life problems across various fields, especially in delivery systems such as mailing, logistics, transportation and product distribution.

\begin{figure}[h!]
\centering
\includegraphics[width=130mm]{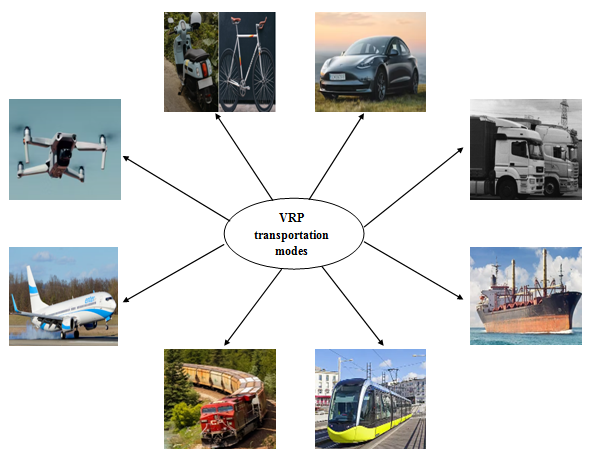}
\caption{Transportation modes in VRP}\label{Fig0}
\end{figure}

In recent years, researchers have sought to make the problem more applicable to real-life scenarios by incorporating additional considerations such as heterogeneous vehicles, multiple depots and customer priority, to name just a few (see Fig. \ref{Fig0}). This diversity has given rise to various variants, such as VRP with time windows, multi-depot VRP, Periodic VRP, Capacitated VRP and others.

\begin{figure}[h!]
\centering
\includegraphics[width=140mm]{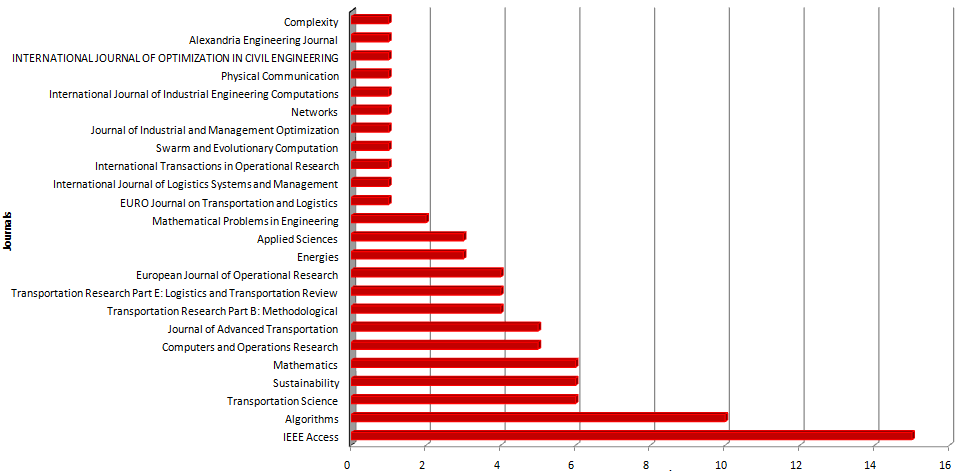}
\caption {Distribution of Selected Articles Across Journals \label{Fig1}}
\end{figure}

\begin{figure}[h!]
\centering
\includegraphics[width=119mm]{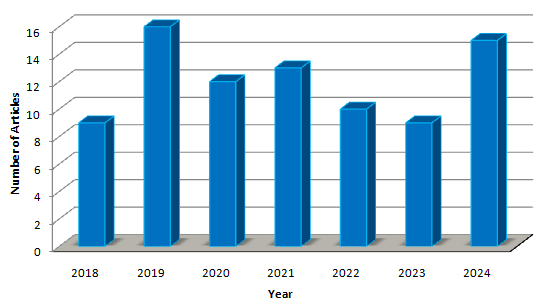} 
\caption{The number of articles selected per Year}
\label{Fig2}
\end{figure}

This article aims to provide a comprehensive overview of the VRP, spanning its fundamental concepts, constraints, objectives, and variants. In order to provide a deeper understanding of the problem and potentially inspire further research in this domain. To achieve this, we reviewed a substantial body of recent literature on the VRP. Our research focused on articles retrieved from the Scopus database (see Fig. \ref{Fig1}), covering papers published between $2018$ and $2024$ (Fig. \ref{Fig2})  depicts the distribution of the selected papers according to the year of publication.

Table \ref{Tab1} provides a comparative overview of our study and existing VRP survey papers. It highlights key aspects such as considered constraints, VRP variants, and optimization objectives. Additionally, it examines the inclusion of recent VRP works, constraint classifications, and whether these surveys tackle research trends and gaps to guide future studies.\\

The remaining of this paper is organized as follows: Section \ref{section2} describes the basic Vehicle Routing Problem, followed by its mathematical formulation in Section \ref{section3}. Section \ref{section4} discusses the basic limitations of the classical VRP. Section \ref{section5} and \ref{section6} address the constraints and variants inherent to the VRP, respectively. Section \ref{section7} outlines the main VRP objectives. Section \ref{section8} examines the current trends and potential future developments in VRP research. The final section presents the conclusion.

\begin{table}[h!]
\caption{Comparison of Vehicle Routing Problem survey papers}\label{Tab1}
\begin{tabular*}{\textwidth}{@{\extracolsep\fill}p{0.5cm} p{1.7cm} p{1.3cm} p{1.5cm} p{1.4cm} p{1.7cm} p{1.5cm} p{6cm}}
\toprule%
\textbf{Ref} & \textbf{Constr.} & \textbf{Variants} & \textbf{Objectives} & \textbf{Scope} & \textbf{Constr. classif.} & \textbf{Trends and gaps} & \textbf{Description} \\
\midrule
\cite{Cao2017} & Yes & No & Limited & 2007-2014 & No & No & Provides a classification of various VRP problems, highlighting common constraints and resolution techniques. \\
\addlinespace
\cite{Muralidharan2018} & Limited & Yes & Limited & 2011-2016 & No & No & Discusses VRP classification and emphasizes bio-inspired solution algorithms.\\
\addlinespace
\cite{Han2018} & Limited & Limited & Yes & 2011-2018 & No & Limited & Classifies and analyzes different VRP variants while reviewing relevant solution methodologies. \\
\addlinespace
\cite{Asghari2020} & Limited & Limited & Yes & 1976-2019 & No & Limited & Reviews recent advances in VRP, focusing on electric VRP, green VRP, and hybrid VRP variants. \\
\addlinespace
\cite{Mor2020} & Limited & Limited & Yes & 1959-2021 & No & Limited & Reviews and classifies the literature on VRPs with temporal decision-making aspects, such as determining the optimal departure time for routes.  \\
\addlinespace
\cite{Injac2024} & Yes & Yes & Limited & 1975-2013 & No & No & Examines various optimization criteria for VRP, including the number of vehicles, traveled distance, and waiting time. It classifies VRP approaches based on their optimization functions. \\
\addlinespace
$[$This$]$ & Yes & Yes & Yes & 2018-2024 & Yes & Yes & Highlights various VRP constraints, objectives, and variants studied in recent research.\\
\botrule
\end{tabular*}
\end{table}

\section{The basic Vehicle Routing Problem} \label{section2}
\begin{figure}[h!]
\centering
\includegraphics[width=119mm]{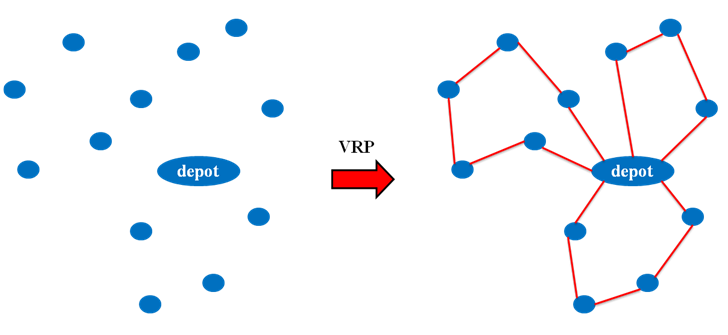}
\caption { Vehicle Routing Problem example \label{Fig3}}
\end{figure}

The vehicle routing problem is, as we mentioned earlier, seeks to determining a set of least-cost routes for a fleet of homogeneous vehicles. The goal is to efficiently serve goods departing from a single depot to a predefined set of customers \ref{Fig3}. This delivery process must adhere to the following constraints \cite{Purnamasari2018, Tan2021}:

\begin{enumerate}[1.]
\item The routing for each vehicle must initiate and conclude at the depot.

  \item The depot itself has no service request.
  
  \item Each customer is served exactly once and by only one vehicle.
  
  \item The total demand along each route should not exceed the capacity of the corresponding vehicle.
  
\end{enumerate}

\section{Basic VRP formulation} \label{section3}
Numerous formulations have been proposed to address the basic Vehicle Routing Problem (BVRP) such as those discussed in \cite{ozaydin2003, Djebbar2021}. In this literature, BVRP is usually modeled using a graph $G=(V,A)$ where:

\begin{itemize}
    \item $V =\{v_0, v_1, \ldots, v_n\}$ represents a set of nodes, where \( v_0 \) designates the central depot, and \( V \setminus \{v_0\} \) corresponds to customers. 
    \item $A = \{(v_i, v_j) \,|\, v_i, v_j \in V \text{ and } v_i \neq v_j\}$ is the set of edges connecting each pair of vertices in $V$, representing the direct routes between customers $i$ and $j$.
\end{itemize}

\begin{figure}[h!]
\centering
\includegraphics[width=119mm]{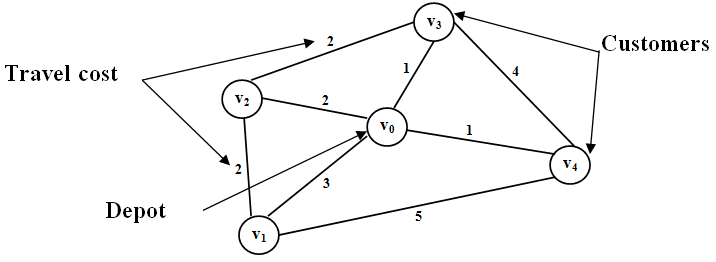}
\caption { VRP graph instance \label{Fig4}}
\end{figure}

Fig. \ref{Fig4} gives an example of such a graph. 

By refering to the graph $G$, modeling VRP by a mathematical program requires defining the following additional parameters:
\begin{itemize}
    \item $K$: the number of available vehicles;
    \item $Q$: the maximum capacity of each vehicle.
    \item $c_{ij}$: the cost incurred for traveling between nodes $v_i$ and $v_j$.
    \item $ q_i$: demand of customer $i$.
\end{itemize}
To define the solutions, BVRP uses the following decision variable:\\
\[
{x_{ijk}} =
\begin{cases}
    1& \text{if vehicle }  k \text{ travels from node } v_i \text{ to node } v_j, \\
    0 & \text{otherwise}.
\end{cases}
\]
Building upon this decision variable, the model can be developed as follows:

\begin{equation}\label{eq:equa1}
\min \sum_{i=1}^{n} \sum_{j=1}^{n} \sum_{k=1}^{K} c_{ij} \cdot x_{ijk}
\end{equation}
\begin{equation} \label{eq:equa2}
    \sum_{i=1}^{n} d_{i} \cdot \sum_{j=1}^{n} x_{ijk} \leq Q  \qquad \forall k \in \{1, \dots, K\}
\end{equation}
\begin{equation}\label{eq:equa3}
    \sum_{i=1}^{n} \sum_{k=1}^{k} x_{ijk} = 1 \qquad \forall j \in \{1, \dots, n\}
\end{equation}
\begin{equation} \label{eq:equa4}
    \sum_{j=1}^{n} \sum_{k=1}^{k} x_{ijk} = 1 \qquad \forall i \in \{1, \dots, n\}
\end{equation}
\begin{equation} \label{eq:equa5}
    \sum_{i=1}^{n} \sum_{l=1}^{n} x_{ilk}  =  \sum_{l=1}^{n}  \sum_{j=1}^{n} x_{ljk} \quad  \forall k \in \{1, \dots, K\}
\end{equation}
\begin{equation} \label{eq:equa6}
    \sum_{j=1}^{n} x_{0jk} = 1 \qquad \forall k \in \{1, \dots, K\}
\end{equation}
\begin{equation} \label{eq:equa7}
    \sum_{i=1}^{n} x_{i0k} = 1 \qquad \forall k \in \{1, \dots, K\}
\end{equation}
\begin{equation} \label{eq:equa8}
   x_{ijk} \in \{0, 1\} 
\end{equation}

In this model, the objective function (\ref{eq:equa1}) optimizes the total cost incurred by the overall vehicles of the fleet. Equation (\ref{eq:equa2}) enforces the capacity constraint by ensuring that the total demand served by a vehicle does not exceed its capacity $Q$. Equations (\ref{eq:equa3}) and (\ref{eq:equa4}) ensure that each customer is visited exactly once, and that each node is leaved exactly once by each vehicle. Equation (\ref{eq:equa5}) ensures the conservation of flow, meaning that every vehicle entering a node must also leave it. Equation (\ref{eq:equa6}) and (\ref{eq:equa7}) ensure that each vehicle starts and ends its route at the depot. Finally, equation (\ref{eq:equa8}) addresses the binary nature of the decision variable.

\section{BVRP limitations} \label{section4}
While BVRP provides a simplified model that is widely used for theoretical analysis, it often falls short in addressing the complexities of real-world applications due to several limitations, including:

\begin{itemize}
    \item Vehicle characteristics: The basic model employs identical and homogeneous vehicles in the delivery process. However, real-world applications often necessitate incorporating vehicles with varying characteristics, such as volume, speed, and capacity.
     
    \item Lack of pickup and delivery considerations: The BVRP primarily addresses the distribution of goods from a depot to customers, excluding scenarios involving pickup and delivery tasks. In many real-life situations, customers may require both pickup and delivery services simultaneously, or may only need items to be picked up from a specific location. The basic VRP omission of these tasks limits its applicability in contexts where integrated pickup and delivery operations are essential.
    
    \item Single visit constraint: In the BVRP, each customer is usually visited once by a single vehicle. This assumption may not hold in cases where the customer's needs exceed the capacity of one vehicle. In such situations, multiple visits by either the same or different vehicles may be required to fulfill the customer's requests.  
     
    \item Static demands and customer needs: In BVRP, customer demands are assumed to be known in advance and remain static throughout the planning horizon. However, in practice, demand can fluctuate over time, with new requests emerging and existing ones changing or disappearing. This dynamic nature of demand requires more flexible models that can accommodate changes and adapt to the evolving requirements of customers.
    
    \item Fixed travel times: The time taken to move between locations is assumed to be fixed and known in advance. However, actually, travel times can vary due to factors such as traffic congestion, weather conditions, and route closures.   
     
    \item Single depot assumption: The basic model involves a single depot from which all vehicles start and end their routes. This does not reflect real-world situations where multiple depots or distribution centers exist.
    
    \item Ignoring customer availability: Respecting delivery time window increases customer satisfaction. However, BVRP doesn't take into account such considerations.
    
    \item Single objective optimization: BVRP focuses on optimizing a single objective, such as minimizing total distance or overall cost. In reality, there are many additional objectives that must be considered such as the number of vehicles and customer satisfaction. Given that the criteria are often contradictory, optimizing only one objective can lead to the deterioration of the others if they are not taken into account. 
    
\end{itemize}

\section{ VRP extensions } \label{section5}
 When formulating VRP, various extensions are introduced to address real-world complexities. These additional considerations are essential to ensure effective problem-solving. This section provides a classification of the most commonly encountered extensions found in recent literature.

\subsection{Customer demands related considerations}
These extensions are directly linked to the specific requests made by customers and include considerations such as:
\begin{itemize}
    \item Deterministic static demands: Refers to scenarios where customer requests are known and fixed prior to the start of the delivery process. All customer demands are predetermined and communicated to the delivery personnel before operations begin, allowing for precise planning and routing based on fixed information. Such a consideration is adopted by the vast majority of papers, such as \cite{Verma2018, Archetti2018, Cassettari2018, Ahmed2018, Linfati2018, Zhang2018, Bortfeldt2019,Nepomuceno2019, Ali2019, Said2019, Zhao2019, Bianchessi2019, Rothenb2019, Pelletier2019, Andelmin2019, Gschwind2019, Cortes-Murcia2019, Henrique2019, Xing2020, Khoo2020, Maaike2020, Valeria2020, Grigorios2020, Utama2020, Mao2020, Zhang2020, Neves-Moreira2020, Zheng2020,  Baris2020, Paula2021, Maximo2021, Lespay2021, Chen2021, Meiling2021, Penglin2021, Hyunpae2021, Marco2021, Zhu2021, Liu2021, Dengkai2021, Inmaculada2022,  Stavropoulou2022, DannyGarcia2022, Zhuang2022, AmineMasmoud2022, Ahmed2022, Payakorn2022, Guillaume2022, Lilian2022, Vincent2022, Bruglieri2023, Maryam2023, Duan2023, Andres2023, Yuxin2023, Peng2023, Huo2023, Qinge2023, Torkzaban2024, Baptista2024, Vincent2024, Kim2024, Zhang2024, Zhou2024}
    
    \item Deterministic dynamic demands: Refers to scenarios where customer requests can change or evolve during the distribution process. Unlike static demands, these demands are not fixed and may vary over time. \cite{Arnau2018, QIU2018, Bernardo2018, Zheng2019, Rim2024, Xie2024}
    
    \item Stochastic demands: In contrast to deterministic demands, stochastic demands involve uncertainty. Here, customer requests are not known in advance but follow a probabilistic distribution. This means that the exact demands can vary and are characterized by uncertainty, requiring flexible and adaptive routing strategies to accommodate potential variations in customer needs. Authors of \cite{Bernardo2018, Munari2019, Guoming2020} adopt such an approach.
    
\end{itemize}

\subsection{Service assumptions}
These considerations pertain to the manner in which services are provided to customers.
\begin{itemize}
    \item Delivery only: Means that the service involves only delivering goods from a depot to customers, with no pickup operations included \cite{Verma2018, Arnau2018, Archetti2018, Cassettari2018, Ahmed2018, Linfati2018, Munari2019, Said2019, Rothenb2019, Pelletier2019, Henrique2019, Xing2020, Khoo2020, Grigorios2020, Utama2020, Mao2020, Neves-Moreira2020, Lespay2021, Chen2021, Meiling2021, Penglin2021, Jalel2021, DannyGarcia2022, Zhuang2022, AmineMasmoud2022, Ahmed2022, Maryam2023, Duan2023, Peng2023, Maleki2019, Ma2024, Pilati2024, Xiao2024, Zhou2024}.
    
    \item Pickup and delivery: Requires the service to include both the collection of goods from specified origins (pickups) and their subsequent delivery to designated destinations \cite{QIU2018, Nepomuceno2019, Zhao2019, Baris2020, Liu2021, Vincent2022, Yuxin2023, Huo2023, Qinge2023, Vincent2024, Xie2024, Francesco2024}.
    
    \item Split delivery: Fulfilling customer requests necessitates carrying out the delivery operation through multiple separate tours. This might be due to customer demands exceeding the limited vehicle capacity \cite{ QIU2018, Bianchessi2019, Bortfeldt2019, Gschwind2019, Xing2020, Hyunpae2021, Guillaume2022, Yuxin2023, Torkzaban2024, Ma2024}.
    
\end{itemize}

\subsection{Vehicle related extensions}
Encompass various considerations or restrictions associated with the fleet of vehicles used in the routing process.

\begin{itemize}
   \item Vehicle capacity constraint: involve the utilization of a fleet of vehicles, each characterized by a finite loading capacity. In adherence to these constraints, it is imperative that the cumulative quantity delivered during one trip does not exceed the designated capacity \cite{ Verma2018, Archetti2018, Cassettari2018, Ahmed2018, Zhang2018, Bernardo2018, Linfati2018, Arnau2018, QIU2018, Ali2019, Said2019, Bianchessi2019, Rothenb2019, Pelletier2019, Cortes-Murcia2019, Henrique2019, Munari2019, Nepomuceno2019, Zhao2019, Gschwind2019, Zheng2019, Maleki2019, Zheng19, Khoo2020, Maaike2020, Valeria2020, Utama2020, Mao2020, Neves-Moreira2020, Zheng2020, Guoming2020, Xing2020, Grigorios2020, Zhang2020, Baris2020, Maximo2021, Lespay2021,Chen2021, Meiling2021, Penglin2021, Hyunpae2021, Marco2021, Liu2021, Dengkai2021,Paula2021, Jalel2021,  DannyGarcia2022, Guillaume2022, Lilian2022, Vincent2022, Stavropoulou2022, Zhuang2022, AmineMasmoud2022, Duan2023, Yuxin2023, Peng2023, Huo2023, Bruglieri2023, Maryam2023, Andres2023, Qinge2023, Zakir2023, Torkzaban2024, Baptista2024, Vincent2024, Kim2024, Zhang2024, Rim2024, Ma2024, Pilati2024, Xiao2024, Xie2024, Francesco2024, Abid2024, Gasset2024, Jasim2024, Zhou2024}.
   
   \item Limited vehicle number: constitute restrictions on the available quantity of vehicles allocated for a given delivery operation \cite{Verma2018, Archetti2018, Cassettari2018, Zhang2018, Linfati2018, Arnau2018, QIU2018, Rothenb2019, Pelletier2019, Cortes-Murcia2019, Henrique2019, Nepomuceno2019, Zhao2019, Gschwind2019, Zheng2019, Neves-Moreira2020, Zheng2020, Guoming2020, Baris2020, Maximo2021, Chen2021, Penglin2021, Hyunpae2021, Liu2021, Dengkai2021, Zhu2021, Inmaculada2022, Guillaume2022, Lilian2022, Stavropoulou2022, Zhuang2022, Ahmed2022, Huo2023, Andres2023, Qinge2023, Zakir2023, Torkzaban2024, Baptista2024, Vincent2024, Kim2024, Zhang2024, Rim2024, Abid2024}.
   \item Heterogeneity vs homogeneity: Refers to the characteristics of the vehicle fleet used in the delivery operation. Vehicle heterogeneity involves using a fleet of vehicles with distinct attributes such as speed, type, volume, and capacity like in \cite{Nepomuceno2019, Zhao2019, Zheng2019, Maleki2019, Zheng19, Neves-Moreira2020, Zheng2020, Baris2020, Penglin2021, Hyunpae2021, Marco2021, Dengkai2021, Jalel2021, Felix2021, DannyGarcia2022, Lilian2022, Stavropoulou2022, Yuxin2023, Maryam2023, Zakir2023, Torkzaban2024, Xiao2024, Abid2024, Gasset2024, Jasim2024}. Conversely, homogeneity involves using a fleet of vehicles with similar attributes for the delivery operation like in  \cite{Verma2018, Archetti2018, Cassettari2018, Ahmed2018, Zhang2018, Bernardo2018, Linfati2018, Arnau2018,QIU2018, Ali2019, Said2019, Bianchessi2019, Rothenb2019, Pelletier2019, Cortes-Murcia2019, Henrique2019, Munari2019, Bortfeldt2019, Andelmin2019, Gschwind2019, Khoo2020, Maaike2020,Valeria2020, Utama2020, Mao2020, Guoming2020, Xing2020, Grigorios2020, Zhang2020, Maximo2021, Lespay2021, Chen2021, Meiling2021, Liu2021, Paula2021, Zhu2021,Inmaculada2022, Guillaume2022, Vincent2022, Zhuang2022, AmineMasmoud2022, Ahmed2022, Payakorn2022, Duan2023, Peng2023, Huo2023, Bruglieri2023, Andres2023, Qinge2023, Baptista2024, Vincent2024, Kim2024, Zhang2024, Rim2024, Ma2024, Pilati2024, Xie2024, Francesco2024, Zhou2024}  
   \item Vehicle type constraints:  Identify the categories of vehicles used in the distribution process, such as electric vehicles \cite{Verma2018,Pelletier2019, Cortes-Murcia2019, Zhao2019, Mao2020, Zhu2021, DannyGarcia2022, AmineMasmoud2022, Duan2023, Bruglieri2023, Maryam2023, Andres2023, Kim2024, Abid2024}, or drones \cite{Zheng19, Jalel2021, Felix2021, Xiao2024, Jasim2024, Zhou2024}.  
   \item Vehicle return constraints: Specifies the post-distribution trajectory of vehicles. Certain configurations mandate vehicles to return directly to the depot at the end of the process, like in \cite{ Verma2018, Archetti2018, Cassettari2018, Ahmed2018, Zhang2018, Bernardo2018, Linfati2018, Arnau2018, QIU2018, Ali2019, Said2019, Bianchessi2019, Rothenb2019, Pelletier2019, Cortes-Murcia2019, Henrique2019, Munari2019, Bortfeldt2019, Nepomuceno2019, Zhao2019, Andelmin2019, Gschwind2019, Zheng2019, Zheng19, Khoo2020, Utama2020, Mao2020, Neves-Moreira2020, Zheng2020, Guoming2020, Xing2020, Grigorios2020, Zhang2020, Baris2020, Maximo2021, Lespay2021,Chen2021, Meiling2021, Penglin2021, Hyunpae2021, Marco2021, Liu2021, Dengkai2021,Paula2021, Zhu2021, Jalel2021, Felix2021, Inmaculada2022, DannyGarcia2022, Guillaume2022, Lilian2022, Vincent2022, Stavropoulou2022, Zhuang2022, AmineMasmoud2022, Ahmed2022, Payakorn2022, Duan2023, Yuxin2023, Peng2023, Huo2023, Bruglieri2023, Maryam2023, Andres2023, Qinge2023, Torkzaban2024, Baptista2024, Vincent2024, Kim2024, Zhang2024, Rim2024, Ma2024, Pilati2024, Xiao2024, Xie2024, Francesco2024, Abid2024, Jasim2024, Zhou2024}, while alternative configurations permit vehicles to be assigned to diverse locations, like in \cite{ Maleki2019, Valeria2020, Zakir2023, Gasset2024}.
\end{itemize}

\subsection{Time related extensions} 
These considerations are associated with the time required to service customers.
\begin{itemize}
   \item Deterministic time windows: Specific time intervals during which goods can be delivered to customers. Each customer is assigned a unique time window, defined before initiating the delivery operation. It also refers to a specific time intervals during which activities at the depot are allowed \cite{Verma2018, Zhang2018, Said2019, Bianchessi2019, Rothenb2019, Cortes-Murcia2019, Munari2019, Zhao2019, Khoo2020, Maaike2020, Mao2020, Zheng2020, Guoming2020, Grigorios2020, Lespay2021, Chen2021, Meiling2021, Liu2021, Dengkai2021, Paula2021, Zhu2021, DannyGarcia2022, Guillaume2022, AmineMasmoud2022, Payakorn2022, Duan2023, Peng2023, Bruglieri2023, Maryam2023, Qinge2023, Zakir2023, Baptista2024, Vincent2024, Kim2024, Zhang2024, Xie2024, Zhou2024}.
   \item Demand periodicity: Recurrence of customer demands at regular intervals. For instance, a customer may submit a request with the intention of replicating it on a daily, weekly, monthly, or other periodic basis \cite{Archetti2018, Rothenb2019, Neves-Moreira2020, Lespay2021, Hyunpae2021, Inmaculada2022, Stavropoulou2022}.
   \item Stochastic servicing time: The time required to service each customer is variable and subject to uncertainty \cite{Munari2019, Guoming2020}.
\end{itemize}
\subsection{Depot related considerations}
They are restrictions associated with the number of depots to be considered.
\begin{itemize}
   \item Single vs multiple depot: Customer deliveries can originate from either a single designated depot or from multiple depots. The single depot constraint requires all deliveries to start from one location \cite{ Verma2018, Archetti2018, Ahmed2018, Zhang2018, Bernardo2018, Linfati2018, Arnau2018, QIU2018, Ali2019, Said2019, Bianchessi2019, Rothenb2019, Pelletier2019, Cortes-Murcia2019, Henrique2019, Munari2019, Bortfeldt2019, Nepomuceno2019, Zhao2019, Andelmin2019, Gschwind2019, Zheng2019, Maleki2019, Zheng19, Khoo2020, Maaike2020, Valeria2020, Utama2020, Mao2020, Neves-Moreira2020, Zheng2020, Guoming2020, Xing2020, Grigorios2020, Baris2020, Maximo2021, Lespay2021,Chen2021, Meiling2021, Penglin2021, Marco2021, Liu2021,Zhu2021, Jalel2021, Felix2021, Inmaculada2022, DannyGarcia2022, Lilian2022, Vincent2022, Stavropoulou2022, Zhuang2022, AmineMasmoud2022, Payakorn2022, Duan2023, Yuxin2023, Peng2023, Huo2023, Bruglieri2023, Maryam2023, Qinge2023, Zakir2023, Torkzaban2024, Baptista2024, Kim2024, Zhang2024, Rim2024, Ma2024, Pilati2024, Xiao2024, Xie2024, Francesco2024, Abid2024, Gasset2024, Jasim2024, Zhou2024}, while the multiple depot assumptions allows for deliveries from various depots, offering greater flexibility and efficiency in routing and scheduling \cite{Cassettari2018, Zhang2020, Hyunpae2021, Dengkai2021, Paula2021, Guillaume2022, Ahmed2022, Andres2023, Vincent2024}.
\end{itemize}

Table \ref{Tab2} recapitulates the extensions usage over the last years.

\begin{table}[h!]
\caption{VRP constraints}\label{Tab2}
\begin{tabular*}{\textwidth}{@{\extracolsep\fill} c p{0.55cm}p{0.55cm}p{0.55cm}p{0.55cm}p{0.55cm}p{0.55cm}p{0.55cm}p{0.55cm}p{0.55cm}p{0.55cm}p{0.55cm}p{0.55cm}p{0.55cm}p{0.55cm}p{0.55cm}p{0.55cm}p{0.55cm}p{0.55cm}}
\toprule
\multirow{5}{*} {\centering \small Ref.} 
& \multicolumn{3}{@{}c@{}}{\centering{\small Customer}} 
& \multicolumn{3}{@{}c@{}}{\centering{ \small Service}} 
& \multicolumn{7}{@{}c@{}}{\centering {\small Vehicle}} 
& \multicolumn{3}{@{}c@{}}{\centering {\small Time}} 
& \multicolumn{2}{@{}c@{}}{\centering {\small Depot}} \\
\cmidrule{2-4} \cmidrule{5-7} \cmidrule{8-14} \cmidrule{15-17} \cmidrule{18-19}
&\multirow{2}{*}{\rotatebox{60}{{\centering \small DSD}}} 
& \multirow{2}{*}{\rotatebox{60}{{\centering \small DDD}}}
& \multirow{2}{*}{\rotatebox{60}{{\centering \small SD}}} 
& \multirow{2}{*}{\rotatebox{60}{{\centering \small DO}}} 
& \multirow{2}{*}{\rotatebox{60}{{\centering \small PD}}} 
& \multirow{2}{*}{\rotatebox{60}{{\centering \small SplD}}} 
& \multirow{2}{*}{\rotatebox{60}{{\centering \small VC}}} 
& \multirow{2}{*}{\rotatebox{60}{{\centering \small LVN}}}
& \multirow{2}{*}{\rotatebox{60}{{\centering \small Hete}}}
& \multicolumn{2}{@{}c@{}}{\centering {\small Type}} 
& \multicolumn{2}{@{}c@{}}{\centering {\small Return}}         
& \multirow{2}{*}{\rotatebox{60}{{\centering \small DTW}}} 
& \multirow{2}{*}{\rotatebox{60}{{\centering \small DP}}} 
& \multirow{2}{*}{\rotatebox{60}{{\centering \small SST}}} 
& \multirow{2}{*}{\rotatebox{60}{{\centering \small 1-D}}} 
& \multirow{2}{*}{\rotatebox{60}{{\centering \small MD}}} \\
\cmidrule{11-12} \cmidrule{13-14}
& &  & & & & & & & & \rotatebox{60}{\small D} & \rotatebox{60}{\small EV} & \rotatebox{60}{\small RD} & \rotatebox{60}{\small OR} & & &  & & \\
\midrule
\rowcolor{lightgray}
\multicolumn{19}{c}{2018} \\
\cite{Verma2018} & \checkmark &  &  & \checkmark &  &  & \checkmark & \checkmark &  &   & \checkmark & \checkmark &  & \checkmark &  &    & \checkmark & \\
\cite{Archetti2018} & \checkmark &  &  & \checkmark &  &  & \checkmark & \checkmark &  &   &  & \checkmark &  &  & \checkmark &   & \checkmark & \\
       \cite{Cassettari2018} & \checkmark &  &   & \checkmark &  &  & \checkmark & \checkmark &  &   &  & \checkmark &  &  &  &   &  & \checkmark \\
       \cite{Ahmed2018} & \checkmark &  &  & \checkmark &  &  & \checkmark &  &  &   &  & \checkmark &  &  &  &   & \checkmark & \\
       \cite{Zhang2018} & \checkmark &  &   &  &  &  & \checkmark & \checkmark &  &   &  & \checkmark &  & \checkmark &  &   & \checkmark & \\
       \cite{Bernardo2018} &  & \checkmark & \checkmark &  &  &  & \checkmark &  &  &   &  & \checkmark &  &  &  &    & \checkmark & \\
       \cite{Linfati2018} & \checkmark &  &   & \checkmark &  &  & \checkmark & \checkmark &  &   &  & \checkmark &  &  &  &    & \checkmark & \\
       \cite{Arnau2018} &  & \checkmark &  & \checkmark &  &  & \checkmark & \checkmark &  &   &  & \checkmark &  &  &  &    & \checkmark & \\
       \cite{QIU2018} &  & \checkmark &  &  & \checkmark & \checkmark & \checkmark & \checkmark &  &   &  & \checkmark &  &  &  &    & \checkmark & \\             
       \rowcolor{lightgray}
       \multicolumn{19}{c}{2019}\\
       \cite{Ali2019} & \checkmark &  &   &  &  &  & \checkmark &  &  &   &  & \checkmark &  &  &  &    & \checkmark & \\
       \cite{Said2019} & \checkmark &   &  & \checkmark &  &  & \checkmark &  &  &   &  & \checkmark &  & \checkmark &   &  & \checkmark & \\
       \cite{Bianchessi2019} & \checkmark &  &   &  &  & \checkmark & \checkmark &  &  &   &  & \checkmark &  & \checkmark &  &    & \checkmark & \\
       \cite{Rothenb2019} & \checkmark &  &   & \checkmark &  &  & \checkmark & \checkmark &  &   &  & \checkmark &  & \checkmark & \checkmark &   & \checkmark & \\
       \cite{Pelletier2019} & \checkmark &  &   & \checkmark &  &  & \checkmark & \checkmark &  &   & \checkmark & \checkmark &  &  &  &    & \checkmark & \\
       \cite{Cortes-Murcia2019} & \checkmark &  &  &  &  &  & \checkmark & \checkmark &  &   & \checkmark & \checkmark &  & \checkmark &    &  & \checkmark & \\
       \cite{Henrique2019} & \checkmark &  &   & \checkmark &  &  & \checkmark & \checkmark &  &   &  & \checkmark &  &  &  &    & \checkmark & \\
       \cite{Munari2019} &  &  & \checkmark & \checkmark &  &  & \checkmark &  &  &   &  & \checkmark &  & \checkmark &  & \checkmark   & \checkmark & \\
       \cite{Bortfeldt2019} & \checkmark  &  &   &  &  & \checkmark &  &  &  &   &  & \checkmark &  &  &  &    & \checkmark & \\
       \cite{Nepomuceno2019} & \checkmark &  &   &  & \checkmark &  & \checkmark & \checkmark & \checkmark &    &  & \checkmark &  &  &  &    & \checkmark & \\       
       \cite{Zhao2019} & \checkmark &  &   &  & \checkmark &  & \checkmark & \checkmark & \checkmark &    & \checkmark & \checkmark &  & \checkmark &  &   & \checkmark & \\
       \cite{Andelmin2019} & \checkmark &  &   &  &  &  &  &  &  &   &  & \checkmark &  &  &  &   & \checkmark & \\       
       \cite{Gschwind2019} & \checkmark &  &   &  &  & \checkmark & \checkmark & \checkmark &  &   &  & \checkmark &  &  &  &  & \checkmark & \\
       \cite{Zheng2019} &  & \checkmark &  &  &  &  & \checkmark & \checkmark & \checkmark &    &  & \checkmark &  &  &  &    & \checkmark & \\
       \cite{Maleki2019} &  &  &   & \checkmark &  &  & \checkmark &  & \checkmark &    &  &  & \checkmark &  &  &    & \checkmark & \\
       \cite{Zheng19} &  &  &   &  &  &  & \checkmark &  & \checkmark & \checkmark  &  & \checkmark &  &  &  &    & \checkmark & \\
       \rowcolor{lightgray}
       \multicolumn{19}{c}{2020}\\
       \cite{Khoo2020} & \checkmark &  &  & \checkmark &  &  & \checkmark &  &  &   &  & \checkmark &  & \checkmark &  &   & \checkmark & \\
       \cite{Maaike2020} & \checkmark &  &   &  &  &  & \checkmark &  &  &   &  &  &  & \checkmark &  &   & \checkmark & \\
       \cite{Valeria2020} & \checkmark &  &   &  &  &  & \checkmark &  &  &   &  &  & \checkmark &  &  &    & \checkmark & \\
       \cite{Utama2020} & \checkmark &  &   & \checkmark &  &  & \checkmark &  &  &   &  & \checkmark &  &  &  &   & \checkmark & \\
       \cite{Mao2020} & \checkmark &  &   & \checkmark &  &  & \checkmark &  &  &   & \checkmark & \checkmark &  & \checkmark &  &   & \checkmark & \\
       \cite{Neves-Moreira2020} & \checkmark &  &  & \checkmark &  &  & \checkmark & \checkmark & \checkmark &    &  & \checkmark &  &  & \checkmark &    & \checkmark & \\
       \cite{Zheng2020} & \checkmark &  &   &  &  &  & \checkmark & \checkmark & \checkmark	 &   &  & \checkmark &  & \checkmark &  &    & \checkmark & \\
       \cite{Guoming2020} &  &  & \checkmark &  &  &  & \checkmark & \checkmark &  &   &  & \checkmark &  & \checkmark &  & \checkmark   & \checkmark & \\   
       \cite{Xing2020} & \checkmark &  &   & \checkmark &  & \checkmark & \checkmark &  &  &   &  & \checkmark &  &  &  &    & \checkmark & \\            
       \cite{Grigorios2020} & \checkmark &  &  & \checkmark &  &  & \checkmark &  &  &   &  & \checkmark &  & \checkmark &  &   & \checkmark & \\      
       \cite{Zhang2020} & \checkmark  &  &   &  &  &  & \checkmark &  &  &   &  & \checkmark &  &  &  &    &  & \checkmark \\      
       \cite{Baris2020} & \checkmark &  &  &  & \checkmark &  & \checkmark & \checkmark & \checkmark &    &  & \checkmark &  &  &  &    & \checkmark & \\     
       \rowcolor{lightgray}
       \multicolumn{19}{c}{2021}\\
       \cite{Maximo2021} & \checkmark &  &   &  &  &  & \checkmark & \checkmark &  &    &  & \checkmark &  &  &  &    & \checkmark & \\
       \cite{Lespay2021} & \checkmark &  &   & \checkmark &  &  & \checkmark &  &  &   &  & \checkmark &  & \checkmark & \checkmark &   & \checkmark & \\      
       \cite{Chen2021} & \checkmark &  &   & \checkmark &  &  & \checkmark & \checkmark &  &   &  & \checkmark &  & \checkmark &  &   & \checkmark & \\
       \cite{Meiling2021} & \checkmark &  &  & \checkmark &  &  & \checkmark &  &  &   &  & \checkmark &  & \checkmark &  &    & \checkmark & \\
       \cite{Penglin2021} & \checkmark &  &   & \checkmark &  &  & \checkmark & \checkmark & \checkmark &    &  & \checkmark &  &  &  &    & \checkmark & \\
       \cite{Hyunpae2021} & \checkmark &  &   &  &  & \checkmark & \checkmark & \checkmark & \checkmark &   &  & \checkmark &  &  & \checkmark   & 
        &  & \checkmark \\
       \cite{Marco2021} & \checkmark &  &   &  &  &  & \checkmark &  & \checkmark &    &  & \checkmark &  &  &  &    & \checkmark & \\
       \cite{Liu2021} & \checkmark &  &   &  & \checkmark &  & \checkmark & \checkmark &  &   &  & \checkmark &  & \checkmark &  &   & \checkmark & \\
       \cite{Dengkai2021} & \checkmark &  &   &  &  &  & \checkmark & \checkmark & \checkmark &    &  & \checkmark &  & \checkmark &  &    &  & \checkmark \\       
       \cite{Paula2021} & \checkmark &  &  &  &  &  & \checkmark &  &  &   &  & \checkmark &  & \checkmark &  &    &  & \checkmark \\
       \cite{Zhu2021} & \checkmark &  &   &  &  &  &  & \checkmark &  &   & \checkmark & \checkmark &  & \checkmark &  &    & \checkmark & \\
       \cite{Jalel2021} &  &   &  & \checkmark &  &  & \checkmark &  & \checkmark &   \checkmark &  & \checkmark &  &  &  &    & \checkmark & \\
       \cite{Felix2021} &  &   &  &  &  &  &  &  & \checkmark &   \checkmark &  & \checkmark &  &  &  &    & \checkmark & \\    
       \rowcolor{lightgray}
       \multicolumn{19}{c}{2022}\\
       \cite{Inmaculada2022} & \checkmark &  &   &  &  &  &  & \checkmark &  &   &  & \checkmark &  &  & \checkmark &    & \checkmark & \\
       \cite{DannyGarcia2022} & \checkmark &  &   & \checkmark &  &  & \checkmark &  & \checkmark &    & \checkmark & \checkmark &  & \checkmark &  &    & \checkmark & \\
       \cite{Guillaume2022} & \checkmark &  &   &  &  & \checkmark & \checkmark & \checkmark &  &   &  & \checkmark &  & \checkmark &  &       &  & \checkmark \\
       \cite{Lilian2022} & \checkmark &  &   &  &  &  & \checkmark & \checkmark & \checkmark &    &  & \checkmark &  &  &  &    & \checkmark & \\
       \cite{Vincent2022} & \checkmark &  &   &  & \checkmark &  & \checkmark &  &  &   &  & \checkmark &  &  &  &    & \checkmark & \\
       \cite{Stavropoulou2022} & \checkmark &  &   &  &  &  & \checkmark & \checkmark & \checkmark &   &  & \checkmark &  &  & \checkmark &    & \checkmark & \\
       \cite{Zhuang2022} & \checkmark &  &   & \checkmark &  &  & \checkmark & \checkmark &  &   &  & \checkmark &  &  &  &    & \checkmark & \\
       \cite{AmineMasmoud2022} & \checkmark &  &   & \checkmark &  &  & \checkmark &  &  &   & \checkmark & \checkmark &  & \checkmark &  &    & \checkmark & \\
       \cite{Ahmed2022} & \checkmark &  &   & \checkmark &  &  &  & \checkmark &  &    &  & \checkmark &  &  &  &    &  & \checkmark \\
       \cite{Payakorn2022} & \checkmark &  &   &  &  &  &  &  &  &   &  & \checkmark &  & \checkmark &  &   & \checkmark & \\
       \rowcolor{lightgray}
       \multicolumn{19}{c}{2023}\\
       \cite{Duan2023} & \checkmark &  &   & \checkmark &  &  & \checkmark &  &  &   & \checkmark & \checkmark &  & \checkmark &  &   & \checkmark & \\
       \cite{Yuxin2023} & \checkmark &  &  &  & \checkmark & \checkmark & \checkmark &  & \checkmark &    &  & \checkmark &  &  &  &    & \checkmark & \\
       \cite{Peng2023} & \checkmark &  &   & \checkmark &  &  & \checkmark &  &  &   &  & \checkmark &  & \checkmark &  &   & \checkmark & \\
       \cite{Huo2023} & \checkmark &  &   &  & \checkmark &  & \checkmark & \checkmark &  &   &  & \checkmark &  &  &  &    & \checkmark & \\
       \cite{Bruglieri2023} & \checkmark &  &   &  &  &  & \checkmark &  &  &   & \checkmark & \checkmark &  & \checkmark &  &    & \checkmark & \\
       \cite{Maryam2023} & \checkmark &  &   & \checkmark &  &  & \checkmark &  & \checkmark &    & \checkmark & \checkmark &  & \checkmark &  &  & \checkmark & \\
       \cite{Andres2023} & \checkmark &  &   &  &  &  & \checkmark & \checkmark &  &   & \checkmark & \checkmark &  &  &  &   &  & \checkmark \\
       \cite{Qinge2023} & \checkmark &  &   &  & \checkmark &  & \checkmark & \checkmark &  &   &  & \checkmark &  & \checkmark &  &    & \checkmark & \\
       \cite{Zakir2023} &  &  &    &  &  &  & \checkmark & \checkmark & \checkmark &    &  &  & \checkmark & \checkmark &  &   & \checkmark & \\
       \rowcolor{lightgray}
       \multicolumn{19}{c}{2024}\\
       \cite{Torkzaban2024} & \checkmark &  &   &  &  & \checkmark & \checkmark & \checkmark & \checkmark &    &  & \checkmark &  &  &  &    & \checkmark & \\
       \cite{Baptista2024} & \checkmark &  &   &  &  &  & \checkmark & \checkmark &  &   &  & \checkmark &  & \checkmark &  &   & \checkmark & \\
       \cite{Vincent2024} & \checkmark &  &   &  & \checkmark &  & \checkmark & \checkmark &  &   &  & \checkmark &  & \checkmark &  &   &  & \checkmark \\
       \cite{Kim2024} & \checkmark &  &   &  &  &  & \checkmark & \checkmark &  &   & \checkmark & \checkmark &  & \checkmark &  &    & \checkmark &  \\
       \cite{Zhang2024} & \checkmark &  &   &  &  &  & \checkmark & \checkmark &  &   &  & \checkmark &  & \checkmark &  &   & \checkmark & \\
       \cite{Rim2024} &  & \checkmark &  &  &  &  & \checkmark & \checkmark &  &   &  & \checkmark &  &  &  &   & \checkmark & \\
       \cite{Ma2024} &  &  &   & \checkmark &  & \checkmark & \checkmark &  &  &   &  & \checkmark &  &  &  &    & \checkmark & \\
       \cite{Pilati2024} &  &  &  & \checkmark &  &  & \checkmark &  &  &   &  & \checkmark &  &  &  &    & \checkmark & \\
       \cite{Xiao2024} &  &   &  & \checkmark &  &  & \checkmark &  & \checkmark &   \checkmark &  & \checkmark &  &  &  &    & \checkmark & \\ 
       \cite{Xie2024} &  &  \checkmark  &  &  & \checkmark &  & \checkmark &  &  &   &  & \checkmark &  & \checkmark &  &   & \checkmark & \\ 
       \cite{Francesco2024} &  &    &  &  & \checkmark &  & \checkmark &  &  &   &  & \checkmark &  &  &  &   & \checkmark & \\         
       \cite{Abid2024} &  &  &   &  &  &  & \checkmark & \checkmark & \checkmark &    & \checkmark & \checkmark &  &  &  &    & \checkmark & \\  
       \cite{Gasset2024} &  &    &  &  &  &  & \checkmark &  & \checkmark &   &  &  & \checkmark &  &  &    & \checkmark & \\ 
       \cite{Jasim2024} &  &    &  &  &  &  & \checkmark &  & \checkmark &   \checkmark &  & \checkmark &  &  &  &    & \checkmark & \\ 
       \cite{Zhou2024} & \checkmark &  &  &  \checkmark &  &  &  \checkmark &  & \checkmark & \checkmark &  & \checkmark &  & \checkmark &  &  & \checkmark &  \\
\bottomrule
\end{tabular*}
\end{table}
\parbox{\textwidth}{
\textbf{Abbreviations:} DSD = Deterministic Static Demands,  
DDD = Deterministic Dynamic Demands, SD = Stochastic Demands,  
DO = Delivery Only, PD = Pickup and Delivery, SplD = Split Delivery,  
VC = Vehicle Capacity, LVN = Limited Vehicle Number,  
Hete = Heterogeneous Fleet, D = Drone, EV = Electric Vehicle, RD = Return Depot,  OR = Open Return, DTW = Deterministic Time Windows,  
DP = Demand Periodicity, SST = Stochastic Serving Time,  
1-D = Single Depot, MD = Multiple Depots.}

\section{VRP Variants}\label{section6}
As the basic Vehicle Routing Problem evolved to meet real-world circumstances, the various considered extensions gave rise to multiple variants with quite distinct VRP research sub-communities. This section offers an overview of the main VRP variants.

\subsection{Stochastic Vehicle Routing Problem}
Stochastic Vehicle Routing Problem (SVRP) is class of VRPs that explicitly consider uncertainties and randomness in various aspects of the routing process. Depending on the specific stochastic variable involved \cite{Bernardo2018, Munari2019, Guoming2020} .
\subsection{Dynamic Vehicle Routing Problem}
In the Dynamic Vehicle Routing Problem (DVRP), the complete set of customer requests is not known in advance and continues to evolve throughout the distribution process. Consequently, as new requests emerge during operations, real-time route adjustments are necessary to incorporate these requests effectively and ensure optimal routing \cite{Arnau2018, QIU2018, Bernardo2018, Zheng2019, Rim2024, Xie2024}.
\subsection{Pickup and Delivery Vehicle Routing Problem}
The Pickup and Delivery Vehicle Routing Problem (PDVRP) involves routing vehicles to pick up items or goods from designated pickup locations and deliver them to specified delivery points. The crucial aspect of this problem is that each pickup must be paired with a corresponding delivery, and these delivery points do not necessarily coincide with the original pickup locations. The works presented in \cite{QIU2018, Nepomuceno2019, Zhao2019, Baris2020, Liu2021, Vincent2022, Yuxin2023, Huo2023, Qinge2023, Vincent2024, Xie2024, Francesco2024} pertain to this category.
\subsection{Split Delivery Vehicle Routing Problem}
Unlike the traditional VRP,  where each vehicle serves a customer?s request in a single tour and must not exceed its capacity, the Split Delivery Vehicle Routing Problem (SDVRP) allows customer requests to be split and hence, fulfilled through multiple tours and potentially by different vehicles. Such an approach is adopted by \cite{ QIU2018, Bianchessi2019, Bortfeldt2019, Gschwind2019, Xing2020, Hyunpae2021, Guillaume2022, Yuxin2023, Torkzaban2024, Ma2024}.
\subsection{Vehicle Routing Problem with Backhauls}
The Vehicle Routing Problem with Backhauls pertains to a scenario where we deal with two distinct customer subsets. The first subset necessitates deliveries from the depot; while the second requires the collection of goods for subsequent delivery to the depot. It is essential to ensure that the pickup operation occurs only after the delivery operations have been completed; all while adhering to the total capacity constraint for each vehicle \cite{Liu2021, Vincent2022, Huo2023, Francesco2024}.
\subsection{Capacitated Vehicle Routing Problem }
In Capacitated Vehicle Routing Problem (CVRP), each vehicle involved in the delivery operation has a limited load-carrying capacity. The primary objective of this variant is to optimize the assignment and sequencing of deliveries across the fleet of vehicles while ensuring that the total demand assigned to each vehicle does not exceed its capacity constraint. This involves determining the most efficient routes for the vehicles to service all customers without violating the capacity limitations \cite{ Verma2018, Archetti2018, Cassettari2018, Ahmed2018, Zhang2018, Bernardo2018, Linfati2018, Arnau2018, QIU2018, Ali2019, Said2019, Bianchessi2019, Rothenb2019, Pelletier2019, Cortes-Murcia2019, Henrique2019, Munari2019, Nepomuceno2019, Zhao2019, Gschwind2019, Zheng2019, Maleki2019, Zheng19, Khoo2020, Maaike2020, Valeria2020, Utama2020, Mao2020, Neves-Moreira2020, Zheng2020, Guoming2020, Xing2020, Grigorios2020, Zhang2020, Baris2020, Maximo2021, Lespay2021,Chen2021, Meiling2021, Penglin2021, Hyunpae2021, Marco2021, Liu2021, Dengkai2021,Paula2021, Jalel2021,  DannyGarcia2022, Guillaume2022, Lilian2022, Vincent2022, Stavropoulou2022, Zhuang2022, AmineMasmoud2022, Duan2023, Yuxin2023, Peng2023, Huo2023, Bruglieri2023, Maryam2023, Andres2023, Qinge2023, Zakir2023, Torkzaban2024, Baptista2024, Vincent2024, Kim2024, Zhang2024, Rim2024, Ma2024, Pilati2024, Xiao2024, Xie2024, Francesco2024, Abid2024, Gasset2024, Jasim2024, Zhou2024}.
\subsection{Vehicle Routing Problem with Fixed Fleet Size}
Vehicle Routing Problem with Fixed Fleet Size (VRPFFS) is a VRP category that focuses on determining the optimal routes for servicing a set of customers while adhering to a constraint on the maximum number of vehicles used. \cite{Verma2018, Archetti2018, Cassettari2018, Zhang2018, Linfati2018, Arnau2018, QIU2018, Rothenb2019, Pelletier2019, Cortes-Murcia2019, Henrique2019, Nepomuceno2019, Zhao2019, Gschwind2019, Zheng2019, Neves-Moreira2020, Zheng2020, Guoming2020, Baris2020, Maximo2021, Chen2021, Penglin2021, Hyunpae2021, Liu2021, Dengkai2021, Zhu2021, Inmaculada2022, Guillaume2022, Lilian2022, Stavropoulou2022, Zhuang2022, Ahmed2022, Huo2023, Andres2023, Qinge2023, Zakir2023, Torkzaban2024, Baptista2024, Vincent2024, Kim2024, Zhang2024, Rim2024, Abid2024}. 
\subsection{Heterogeneous Vehicle Routing Problem}
In the BVRP, the vehicles have identical characteristics. However in the Heterogeneous Vehicle Routing Problem (HVRP) such an assumption is alleviated. These characteristics may include differences in capacity, size, volume, speed, and other attributes. The goal is to efficiently deliver goods to a set of customers with diverse requests and locations, taking into account the distinct capabilities of each vehicle in the fleet \cite{Nepomuceno2019, Zhao2019, Zheng2019, Maleki2019, Zheng19, Neves-Moreira2020, Zheng2020, Baris2020, Penglin2021, Hyunpae2021, Marco2021, Dengkai2021, Jalel2021, Felix2021, DannyGarcia2022, Lilian2022, Stavropoulou2022, Yuxin2023, Maryam2023, Zakir2023, Torkzaban2024, Xiao2024, Abid2024, Gasset2024, Jasim2024, Zhou2024}.
\subsection{Vehicle Routing Problem with Drones}
The Vehicle Routing Problem with Drones (VRPD) focuses exclusively on the use of drones for delivering goods. The objective is to determine the optimal routes for a fleet of drones to serve a set of customers, aiming to efficiently manage deliveries by minimizing energy consumption besides classical factors such as travel time \cite{Zheng19, Jalel2021, Felix2021, Xiao2024, Jasim2024, Zhou2024}.
\subsection{Electric Vehicle Routing Problem} 
The Electric Vehicle Routing Problem (EVRP) focuses on servicing customers using electric vehicles. The primary objective of this variant is to optimize the routing and scheduling to minimize the energy consumption while ensuring efficient service delivery \cite{Verma2018,Pelletier2019, Cortes-Murcia2019, Zhao2019, Mao2020, Zhu2021, DannyGarcia2022, AmineMasmoud2022, Duan2023, Bruglieri2023, Maryam2023, Andres2023, Kim2024, Abid2024}.
\subsection{Open Vehicle Routing Problem}
Open Vehicle Routing Problem (OVRP) is a variant of the classical VRP in which vehicles are not required to return to the depot upon completing their deliveries. Instead, each vehicle concludes its route at the last assigned customer or at a specified endpoint \cite{ Maleki2019, Valeria2020, Zakir2023, Gasset2024}.
\subsection{Vehicle Routing Problem with Time Window}
The Vehicle Routing Problem with Time Window (VRPTW) incorporates time constraints into the problem formulation. Each customer is assigned a specific time interval during which he is available to receive his goods or services \cite{Verma2018, Zhang2018, Said2019, Bianchessi2019, Rothenb2019, Cortes-Murcia2019, Munari2019, Zhao2019, Khoo2020, Maaike2020, Mao2020, Zheng2020, Guoming2020, Grigorios2020, Lespay2021, Chen2021, Meiling2021, Liu2021, Dengkai2021, Paula2021, Zhu2021, DannyGarcia2022, Guillaume2022, AmineMasmoud2022, Payakorn2022, Duan2023, Peng2023, Bruglieri2023, Maryam2023, Qinge2023, Zakir2023, Baptista2024, Vincent2024, Kim2024, Zhang2024, Xie2024, Zhou2024}.

There are two type of VRPTW:
\begin{itemize}
   \item Rigid VRPTW: Requires that each customer be served strictly within their specified time window.
   \item Released VRPTW: Allows for some flexibility in timing, permitting deviations from the time windows (The scheduled operations are either delayed or expedited ). However, it is important to note that the deviations from the time windows impose penalties. Therefore, in the released VRPTW, the goal is not only to minimize the total cost but also to ensure that all customers are served within their assigned time windows with least cumulate deviation. 
\end{itemize}
\subsection{Periodic Vehicle Routing Problem} 
Periodic Vehicle Routing Problem (PVRP) Focuses on optimizing routes to service a predefined set of customers requiring deliveries to be made on a regular schedule. For example, a customer may need deliveries every Monday and Friday between 11:00 AM and 2:00 PM. The challenge is to plan and execute repeated delivery operations efficiently with different periodicity time frames \cite{Archetti2018, Rothenb2019, Neves-Moreira2020, Lespay2021, Hyunpae2021, Inmaculada2022, Stavropoulou2022}.
\subsection{Vehicle Routing Problem with Multiple Depots}
The Vehicle Routing Problem with Multiple Depots (VRPMD) involves multiple depots with the vehicles being able to start their routes from any one of them. Depots are often located in various geographical locations. The primary objective is to optimize the routing, ensuring that each vehicle returns to its originating depot upon completing its delivery operations \cite{Cassettari2018, Zhang2020, Hyunpae2021, Dengkai2021, Paula2021, Guillaume2022, Ahmed2022, Andres2023, Vincent2024}.

\subsection{Two-Echelon Vehicle Routing Problem}
The Two-Echelon Vehicle Routing Problem (2E-VRP) focuses on minimizing routing costs across two hierarchical levels. In the first echelon, vehicles transport goods from a central depot to a network of satellite facilities. In the second echelon, additional deliveries are made from these satellite facilities to the final customers. The goal is to optimize the overall routing cost, encompassing both the transportation from the central depot to the satellite facilities and the subsequent distribution to the end customers \cite{Marco2021, Guillaume2022}.
\subsection{Green Vehicle Routing Problem}
The Green Vehicle Routing Problem (GVRP) focuses on optimizing routes and distribution processes while considering environmental factors. The primary objective of this variant is to minimize carbon emissions, fuel consumption, and other environmental impacts associated with transportation \cite{Andelmin2019, Utama2020, Zhang2020}.
\\
\\
\textbf{Remark} Numerous other variants of VRP have been explored in the literature, often involving combinations of two or more of the variants presented above. \\

Table \ref{Tab3} provides an overview of the various VRP variants explored in recent years.

\begin{table}[h!]
\caption{VRP variants}\label{Tab3}
\begin{tabular*}{\textwidth}{@{\extracolsep\fill} m{0.7cm} p{0.7cm}p{0.7cm}p{0.7cm}p{0.7cm}p{0.7cm}p{0.7cm}p{0.7cm}p{0.7cm}p{0.7cm}p{0.7cm}p{0.7cm}p{0.7cm}p{0.7cm}p{0.7cm}p{0.7cm}p{0.7cm}}
\toprule%
{\centering \small{Ref}}   
& \rotatebox{60}{\small SVRP} & \rotatebox{60}{\small DVRP} & \rotatebox{60}{\small PDVRP} & \rotatebox{60}{\small SDVRP} & \rotatebox{60}{\small VRPB} & \rotatebox{60}{\small CVRP} & \rotatebox{60}{\small VRPFFS} & \rotatebox{60}{\small HVRP} & \rotatebox{60}{\small VRPD} & \rotatebox{60}{\small EVRP} & \rotatebox{60}{\small OVRP} & \rotatebox{60}{\small VRPTW} & \rotatebox{60}{\small PVRP} & \rotatebox{60}{\small MDVRP} & \rotatebox{60}{\small 2-EVRP} & \rotatebox{60}{\small GVRP}  \\
\midrule
\rowcolor{lightgray}
    \multicolumn{17}{c}{2018}\\
    \cite{Verma2018} &  &  &  &  &  & \checkmark & \checkmark &  &  & \checkmark &  & \checkmark &  &  &  & \\
    \cite{Archetti2018} &  &  &  &  &  & \checkmark & \checkmark &  &  &  &  &  & \checkmark &  &  & \\
    \cite{Cassettari2018} &  &  &  &  &  & \checkmark & \checkmark &  &   &  &  &  &  & \checkmark &  & \\
    \cite{Ahmed2018} &  &  &  &  &  & \checkmark &  &  &  &  &  &  &  &  &  & \\
    \cite{Zhang2018} &  &  &  &  &  & \checkmark & \checkmark &  &  &  &  & \checkmark &  &  &  & \\
    \cite{Bernardo2018} & \checkmark & \checkmark &  &  &  & \checkmark &  &  &  &  &  &  &  &  &  & \\
    \cite{Linfati2018} &  &  &  &  &  & \checkmark & \checkmark &  &  &  &  &  &  &  &  & \\           
    \cite{Arnau2018} &  & \checkmark &  &  &  & \checkmark & \checkmark &  &  &  &  &  &  &  &  & \\
    \cite{QIU2018} &  & \checkmark & \checkmark & \checkmark &  & \checkmark & \checkmark &  &  &  &  &  &  &  &  & \\   
    \rowcolor{lightgray}
    \multicolumn{17}{c}{2019}\\
    \cite{Ali2019} &  &  &  &  &  & \checkmark &  &  &  &  &  &  &  &  &  & \\
    \cite{Said2019} &  &  &  &  &  & \checkmark &  &  &  &  &  & \checkmark &  &  &  & \\
    \cite{Bianchessi2019} &  &  &  & \checkmark &  & \checkmark &  &  &  &  &  & \checkmark &  &  &  & \\
    \cite{Rothenb2019} &  &  &  &  &  & \checkmark & \checkmark &  &  &  &  & \checkmark & \checkmark &  &  & \\
    \cite{Pelletier2019} &  &  &  &  &  & \checkmark & \checkmark &  &  & \checkmark &  &  &  &  &  & \\
    \cite{Cortes-Murcia2019} &  &  &  &  &  & \checkmark & \checkmark &  &  & \checkmark &  & \checkmark &  &  &  & \\
    \cite{Henrique2019} &  &  &  &  &  & \checkmark & \checkmark &  &  &  &  &  &  &  &  & \\
    \cite{Munari2019} & \checkmark &  &  &  &  & \checkmark &  &  &  &  &  & \checkmark &  &  &  & \\    
    \cite{Bortfeldt2019} &  &  &  & \checkmark &  &  &  &  &  &  &  &  &  &  &  & \\
    \cite{Nepomuceno2019} &  &  & \checkmark &  &  & \checkmark & \checkmark & \checkmark &  &  &  &  &  &  &  & \\
    \cite{Zhao2019} &  &  & \checkmark &  &  & \checkmark & \checkmark & \checkmark &  & \checkmark &  & \checkmark &  &  &  & \\ 
    \cite{Andelmin2019} &  &  &  &  &  &  &  &  &  &  &  &  &  &  &  & \checkmark \\
    \cite{Gschwind2019} &  &  &  & \checkmark &  & \checkmark & \checkmark &  &  &  &  &  &  &  &  & \\
    \cite{Zheng2019} &  & \checkmark &  &  &  & \checkmark & \checkmark & \checkmark &  &  &  &  &  &  &  & \\
    \cite{Maleki2019} &  &  &  &  &  & \checkmark &  & \checkmark &  &  & \checkmark &  &  &  &  & \\
    \cite{Zheng19} &  &  &  &  &  & \checkmark &  & \checkmark & \checkmark &  &  &  &  &  &  & \\
    \rowcolor{lightgray}
    \multicolumn{17}{c}{2020}\\
    \cite{Khoo2020} &  &  &  &  &  & \checkmark &  &  &  &  &  & \checkmark &  &  &  & \\
    \cite{Maaike2020} &  &  &  &  &  & \checkmark &  &  &  &  &  & \checkmark &  &  &  & \\
    \cite{Valeria2020} &  &  &  &  &  & \checkmark &  &  &  &  & \checkmark &  &  &  &  & \\
    \cite{Utama2020} &  &  &  &  &  & \checkmark &  &  &  &  &  &  &  &  &  & \checkmark \\
    \cite{Mao2020} &  &  &  &  &  & \checkmark &  &  &  & \checkmark &  & \checkmark &  &  &  & \\
    \cite{Neves-Moreira2020} &  &  &  &  &  & \checkmark & \checkmark & \checkmark  &  &  &  &  & \checkmark &  &  & \\
    \cite{Zheng2020} &  &  &  &  &  & \checkmark & \checkmark & \checkmark &  &  &  & \checkmark &  &  &  & \\
    \cite{Guoming2020} & \checkmark &  &  &  &  & \checkmark & \checkmark &  &  &  &  & \checkmark &  &  &  & \\
    \cite{Xing2020} &  &  &  & \checkmark &  & \checkmark &  &  &  &  &  &  &  &  &  & \\
    \cite{Grigorios2020} &  &  &  &  &  & \checkmark &  &  &  &  &  & \checkmark &  &  &  & \\
    \cite{Zhang2020} &  &  &  &  &  & \checkmark &  &  &  &  &  &  &  & \checkmark &  & \checkmark \\
    \cite{Baris2020} &  &  & \checkmark &  &  & \checkmark & \checkmark & \checkmark &  &  &  &  &  &  &  & \\
    \rowcolor{lightgray}
    \multicolumn{17}{c}{2021}\\
    \cite{Maximo2021} &  &  &  &  &  & \checkmark & \checkmark &  &  &  &  &  &  &  &  & \\    
    \cite{Lespay2021} &  &  &  &  &  & \checkmark &  &  &  &  &  & \checkmark & \checkmark &  &  & \\  
    \cite{Chen2021} &  &  &  &  &  & \checkmark & \checkmark &  &  &  &  & \checkmark &  &  &  & \\
    \cite{Meiling2021} &  &  &  &  &  & \checkmark &  &  &  &  &  & \checkmark &  &  &  & \\
    \cite{Penglin2021} &  &  &  &  &  & \checkmark & \checkmark & \checkmark &  &  &  &  &  &  &  & \\
    \cite{Hyunpae2021} &  &  &  & \checkmark &  & \checkmark & \checkmark & \checkmark &  &  &  &  & \checkmark & \checkmark &  & \\
    \cite{Marco2021} &  &  &  &  &  & \checkmark &  & \checkmark &  &  &  &  &  &  & \checkmark & \\
    \cite{Liu2021} &  &  & \checkmark &  & \checkmark & \checkmark & \checkmark &  &  &  &  & \checkmark &  &  &  & \\
    \cite{Dengkai2021} &  &  &  &  &  & \checkmark & \checkmark & \checkmark &  &  &  & \checkmark &  & \checkmark &  & \\
    \cite{Paula2021} &  &  &  &  &  & \checkmark &  &  &  &  &  & \checkmark &  & \checkmark &  & \\
    \cite{Zhu2021} &  &  &  &  &  &  & \checkmark &  &  & \checkmark &  & \checkmark &  &  &  & \\
    \cite{Jalel2021} &  &  &  &  &  & \checkmark &  & \checkmark & \checkmark &  &  &  &  &  &  & \\
    \cite{Felix2021} &  &  &  &  &  &  &  & \checkmark & \checkmark &  &  &  &  &  &  & \\  
    \rowcolor{lightgray}
    \multicolumn{17}{c}{2022}\\
    \cite{Inmaculada2022} &  &  &  &  &  &  & \checkmark &  &  &  &  &  & \checkmark &  &  & \\
    \cite{DannyGarcia2022} &  &  &  &  &  & \checkmark &  & \checkmark &  & \checkmark &  & \checkmark &  &  &  & \\
    \cite{Guillaume2022} &  &  &  & \checkmark &  & \checkmark & \checkmark &  &  &  &  & \checkmark &  & \checkmark & \checkmark & \\
    \cite{Lilian2022} &  &  &  &  &  & \checkmark & \checkmark & \checkmark &  &  &  &  &  &  &  & \\
    \cite{Vincent2022} &  &  & \checkmark &  & \checkmark & \checkmark &  &  &  &  &  &  &  &  &  & \\
    \cite{Stavropoulou2022} &  &  &  &  &  & \checkmark & \checkmark & \checkmark &  &  &  &  & \checkmark &  &  & \\
    \cite{Zhuang2022} &  &  &  &  &  & \checkmark & \checkmark &  &  &  &  &  &  &  &  & \\
    \cite{AmineMasmoud2022} &  &  &  &  &  & \checkmark &  &  &  & \checkmark &  & \checkmark &  &  &  & \\
    \cite{Ahmed2022} &  &  &  &  &  &  & \checkmark &  &  &  &  &  &  & \checkmark &  & \\
    \cite{Payakorn2022} &  &  &  &  &  &  &  &  &  &  &  & \checkmark &  &  &  & \\
    \rowcolor{lightgray}
    \multicolumn{17}{c}{2023}\\
    \cite{Duan2023} &  &  &  &  &  & \checkmark &  &  &  & \checkmark &  & \checkmark &  &  &  & \\
    \cite{Yuxin2023} &  &  & \checkmark & \checkmark &  & \checkmark &  & \checkmark &  &  &  &  &  &  &  & \\
    \cite{Peng2023} &  &  &  &  &  & \checkmark &  &  &  &  &  & \checkmark &  &  &  & \\
    \cite{Huo2023} &  &  & \checkmark &  & \checkmark & \checkmark & \checkmark &  &  &  &  &  &  &  &  & \\
    \cite{Bruglieri2023} &  &  &  &  &  & \checkmark &  &  &  & \checkmark &  & \checkmark &  &  &  & \\
    \cite{Maryam2023} &  &  &  &  &  & \checkmark &  & \checkmark &  & \checkmark &  & \checkmark &  &  &  & \\
    \cite{Andres2023} &  &  &  &  &  & \checkmark & \checkmark &  &  & \checkmark &  &  &  & \checkmark &  & \\
    \cite{Qinge2023} &  &  & \checkmark &  &  & \checkmark & \checkmark &  &  &  &  & \checkmark &  &  &  &  \\
    \cite{Zakir2023} &  &  &  &  &  & \checkmark & \checkmark & \checkmark &  &  & \checkmark & \checkmark &  &  &  &  \\
    \rowcolor{lightgray}
    \multicolumn{17}{c}{2024}\\
    \cite{Torkzaban2024} &  &  &  & \checkmark &  & \checkmark & \checkmark & \checkmark &  &  &  &  &  &  &  & \\
    \cite{Baptista2024} &  &  &  &  &  & \checkmark & \checkmark &  &  &  &  & \checkmark &  &  &  & \\
    \cite{Vincent2024} &  &  & \checkmark &  &  & \checkmark & \checkmark &  &  &  &  & \checkmark &  & \checkmark &  & \\
    \cite{Kim2024} &  &  &  &  &  & \checkmark & \checkmark &  &  & \checkmark &  & \checkmark &  &  &  & \\
    \cite{Zhang2024} &  &  &  &  &  & \checkmark & \checkmark &  &  &  &  & \checkmark &  &  &  & \\
    \cite{Rim2024} &  & \checkmark &  &  &  & \checkmark & \checkmark &  &  &  &  &  &  &  &  & \\
    \cite{Ma2024} &  &  &  & \checkmark &  & \checkmark &  &  &  &  &  &  &  &  &  & \\
    \cite{Pilati2024} &  &  &  &  &  & \checkmark &  &  &  &  &  &  &  &  &  & \\
    \cite{Xiao2024} &  &  &  &  &  & \checkmark &  & \checkmark &  \checkmark &  &  &  &  &  &  & \\
    \cite{Xie2024} &  & \checkmark & \checkmark &  &  & \checkmark &  &  &  &  &  & \checkmark &  &  &  & \\
    \cite{Francesco2024} &  &  & \checkmark &  & \checkmark & \checkmark &  &  &  &  &  &  &  &  &  & \\
    \cite{Abid2024} &  &  &  &  &  & \checkmark & \checkmark & \checkmark &  & \checkmark &  &  &  &  &  & \\
    \cite{Gasset2024} &  &  &  &  &  & \checkmark &  & \checkmark &  &  & \checkmark &  &  &  &  & \\
    \cite{Jasim2024} &  &  &  &  &  & \checkmark &  & \checkmark & \checkmark &  &  &  &  &  &  & \\
    \cite{Zhou2024} &  &  &  &  &  & \checkmark &  & \checkmark & \checkmark &  &  & \checkmark &  &  &  & \\
\botrule
\end{tabular*}
\end{table}

\section{VRP objectives} \label{section7}
Based on specific needs and constraints, the literature proposes various objectives. Although minimizing traveling distance remains the primary focus in most related research \cite{Zhu2021}, many VRP papers include numerous other criteria as additional objectives. The following outlines some of them.
\subsection{Total distance minimization}  This objective aims to reduce the total distance covered by the fleet. By optimizing routes to be shorter, this objective mirrors a generalization of the Traveling Salesman Problem within the Vehicle Routing Problem. The works presented in \cite{Verma2018, Cassettari2018, Ahmed2018, Zhang2018, Linfati2018, Arnau2018, QIU2018, Ali2019, Bortfeldt2019,  Andelmin2019, Zheng2019, Maleki2019, Khoo2020, Xing2020, Grigorios2020, Meiling2021, Penglin2021, Hyunpae2021, Liu2021, Dengkai2021, DannyGarcia2022, Guillaume2022, Lilian2022, Zhuang2022, AmineMasmoud2022, Ahmed2022, Duan2023, Andres2023, Zakir2023, Baptista2024, Kim2024, Zhang2024, Rim2024, Ma2024, Francesco2024, Abid2024, Gasset2024, Zhou2024} adopted such an approach.
\subsection{Time minimization} Time minimization aims to reduce the total duration of delivery processes, which is particularly critical in scenarios requiring prompt service, such as the transportation of perishable goods or urgent deliveries. This objective encompasses the reduction of several time-related factors, including:
\begin{itemize}
    \item Travel time: The total duration required for vehicles to complete deliveries to all designated customers. Efficient routing strategies reduce travel time, enhancing overall delivery efficiency \cite{Chen2021, Jalel2021, Felix2021, Qinge2023, Abid2024}.
    \item Waiting time: The time lost due to delays in vehicle charging or recharging processes, as well as non-adherence to predefined time windows. Additionally, waiting time can result from traffic congestion, delays at customer locations due for instance to client unavailability or localization issues.  \cite{Cortes-Murcia2019, Zhao2019, Zheng2019, Maaike2020, Mao2020, Chen2021, Zhu2021, Duan2023}.
\end{itemize}
\subsection{Cost minimization} Cost minimization aims to reduce the expenses associated with the vehicle routing process. This objective covers various cost components, including:
\begin{itemize}
   \item Travel-related expenses: Costs that are directly correlated with distance traveled, such as fuel consumption, energy costs (including battery charging or swapping for drones and electric vehicles) , emissions costs, and maintenance expenses \cite{Verma2018, Archetti2018, Bernardo2018, Said2019, Bianchessi2019, Rothenb2019, Pelletier2019, Henrique2019, Munari2019, Nepomuceno2019, Zhao2019, Gschwind2019, Valeria2020, Utama2020, Mao2020,  Neves-Moreira2020, Guoming2020, Zhang2020, Baris2020, Marco2021, Paula2021, Zhu2021, DannyGarcia2022, Guillaume2022, Lilian2022, Vincent2022, Stavropoulou2022, AmineMasmoud2022, Payakorn2022, Yuxin2023, Peng2023, Bruglieri2023, Maryam2023, Torkzaban2024, Vincent2024, Pilati2024, Xiao2024, Xie2024, Abid2024, Jasim2024, Zhou2024}.
   \item Fixed costs: Costs associated with vehicle deployment, including insurance, vehicle acquisition, and taxes. Reducing these costs through strategic fleet management contributes to overall cost efficiency \cite{Rothenb2019, Pelletier2019, Nepomuceno2019, Zheng19, Utama2020, Mao2020, Baris2020, Maximo2021, Meiling2021, Liu2021, Dengkai2021, Paula2021, Guillaume2022, Lilian2022, Yuxin2023, Bruglieri2023, Maryam2023, Zakir2023, Vincent2024, Xiao2024, Zhou2024}.
\end{itemize}   
\subsection{Minimize the number of vehicles} Minimizing the number of vehicles in delivery operations aims to reduce fleet size while maintaining operational efficiency. This is achieved through optimized route planning and delivery consolidation, ensuring that the minimum necessary number of vehicles is utilized without compromising service quality. Furthermore, reducing fleet size helps to reduce environmental impact and minimize overall vehicle-related costs, including fuel consumption, maintenance, energy costs, discussed in the previous section.  \cite{ Bortfeldt2019, Maleki2019, Valeria2020, Grigorios2020, Lespay2021, DannyGarcia2022, Abid2024}.
\subsection{Minimization of penalties} This objective address the costs associated with service failures, particularly violations of temporal constraints such as time windows and customer priority. Minimizing penalties enhances service reliability and improves customer satisfaction, contributing to more efficient delivery operations \cite{Verma2018, Zheng2020, Guoming2020, Meiling2021, Paula2021, Peng2023, Huo2023, Xie2024, Zhou2024}.
\subsection{Minimization of drivers} This objective aims to reduce the number of drivers needed for deliveries by improving scheduling and route planning. Since drivers work for a limited time and vehicles can be used by multiple drivers, better planning helps keep operations running smoothly with fewer drivers \cite{Paula2021, Zhu2021, Huo2023, Bruglieri2023, Xie2024}. \\
Table \ref{Tab4} provides an overview of the various VRP objectives explored in recent years.

\begin{table}[h]
\caption{VRP objectives}\label{Tab4}
\begin{tabular*}{\textwidth}{@{\extracolsep\fill}>{\centering\arraybackslash}p{1cm}>{\centering\arraybackslash}p{1.6cm}>{\centering\arraybackslash}p{1.6cm}>{\centering\arraybackslash}p{1.6cm}>{\centering\arraybackslash}p{1.6cm}>{\centering\arraybackslash}p{1.6cm}>{\centering\arraybackslash}p{1.6cm}>{\centering\arraybackslash}p{1.6cm}>{\centering\arraybackslash}p{1.6cm}
}
\toprule%
\multirow{8}{*} {\small Ref} & \multirow{8}{*}{\rotatebox{60}{\centering {\centering \small Distance min}}}
    & \multicolumn{2}{@{}c@{}}{\centering{ \small Time min}}   
    & \multicolumn{2}{@{}c@{}}{\centering {\small Cost min}} 
    & \multirow{8}{*}{\rotatebox{60}{\centering {\centering \small Vehicle  min}}}
    & \multirow{8}{*}{\rotatebox{60}{\centering {\centering \small Min penalties}}}
    & \multirow{8}{*}{\rotatebox{60}{\centering {\centering \small Min Drivers}}}\\
    \cmidrule{3-4}\cmidrule{5-6}%
& 
    & {\rotatebox{60}{\centering \small Travel time}}
    & {\rotatebox{60}{\centering \small Waiting time}}
    & {\rotatebox{60}{\centering \small Travel cost}} 
    & {\rotatebox{60}{\centering \small Fixed cost}} 
    & & & \\    
\midrule
\rowcolor{lightgray}
    \multicolumn{9}{c}{2018}\\
    \cite{Verma2018} & \checkmark &  &  & \checkmark &  &  & \checkmark &  \\ 
    \cite{Archetti2018} &  &  &  & \checkmark &  &  &  &\\
    \cite{Cassettari2018} & \checkmark &  &  &  &  &  &  &  \\
    \cite{Ahmed2018} & \checkmark &  &  &  &  &  &  &  \\
    \cite{Zhang2018} & \checkmark &  &  &  &  &  &  &  \\ 
    \cite{Bernardo2018} &  &  &  & \checkmark &  &  &  &   \\
    \cite{Linfati2018} & \checkmark &  &  &  &  &  &  &   \\                  
    \cite{Arnau2018} & \checkmark &  &  &  &  &  &  &  \\
    \cite{QIU2018} & \checkmark &  &  &  &  &  &  &   \\
    \rowcolor{lightgray}
    \multicolumn{9}{c}{2019}\\ 
    \cite{Ali2019} & \checkmark &  &  &  &  &  &  &  \\
    \cite{Said2019} &  &  &  & \checkmark &  &  &  &   \\
    \cite{Bianchessi2019} &  &  &  & \checkmark &  &  &  &   \\
    \cite{Rothenb2019} &  &  &  & \checkmark & \checkmark &  &  &  \\
    \cite{Pelletier2019} &  &  &  & \checkmark & \checkmark &  &  &   \\  
    \cite{Cortes-Murcia2019} &  &  & \checkmark &  &  &  &  &   \\
    \cite{Henrique2019} &  &  &  & \checkmark &  &  &  &   \\
    \cite{Munari2019} &  &  &  & \checkmark &  &  &  &   \\
    \cite{Bortfeldt2019} & \checkmark &  &  &  &  & \checkmark &  &   \\
    \cite{Nepomuceno2019} &  &  &  & \checkmark & \checkmark &  &  &  \\
    \cite{Zhao2019} &  &  & \checkmark & \checkmark &  &  &  &  \\
    \cite{Andelmin2019} & \checkmark &  &  &  &  &  &  &   \\
    \cite{Gschwind2019} &  &  &  & \checkmark &  &  &  &   \\
    \cite{Zheng2019} & \checkmark &  & \checkmark &  &  &  &  &  \\
    \cite{Maleki2019} & \checkmark &  &  &  &  & \checkmark &  & \\
    \cite{Zheng19} &  &  &  &  & \checkmark &  &  &   \\
    \rowcolor{lightgray}
    \multicolumn{9}{c}{2020}\\
    \cite{Khoo2020} & \checkmark &  &  &  &  &  &  &  \\
    \cite{Maaike2020} &  &  & \checkmark &  &  &  &  &   \\
    \cite{Valeria2020} &  &  &  & \checkmark &  & \checkmark &  &  \\
    \cite{Utama2020} &  &  &  & \checkmark & \checkmark &  &  &   \\
    \cite{Mao2020} &  &  & \checkmark & \checkmark & \checkmark &  &  &  \\
    \cite{Neves-Moreira2020} &  &  &  & \checkmark &  &  &  &   \\
    \cite{Zheng2020} &  &  &  &  &  &  & \checkmark &  \\
    \cite{Guoming2020} &  &  &  & \checkmark &  &  & \checkmark &   \\
    \cite{Xing2020} & \checkmark &  &  &  &  &  &  &   \\
    \cite{Grigorios2020} & \checkmark &  &  &  &  & \checkmark &  &   \\
    \cite{Zhang2020} &  &  &  & \checkmark &  &  &  &   \\
    \cite{Baris2020} &  &  &  & \checkmark & \checkmark &  &  &  \\
    \rowcolor{lightgray}
    \multicolumn{9}{c}{2021}\\
    \cite{Maximo2021} &  &  &  &  & \checkmark &  &  &   \\
    \cite{Lespay2021} &  &  &  &  &  & \checkmark &  & \\
    \cite{Chen2021} &  & \checkmark & \checkmark &  &  &  &  &   \\
    \cite{Meiling2021} & \checkmark &  &  &  & \checkmark &  & \checkmark &  \\
    \cite{Penglin2021} & \checkmark &  &  &  &  &  &  &   \\
    \cite{Hyunpae2021} & \checkmark &  &  &  &  &  &  &   \\
    \cite{Marco2021} &  &  &  & \checkmark &  &  &  &   \\
    \cite{Liu2021} & \checkmark &  &  &  & \checkmark &  &  &   \\
    \cite{Dengkai2021} & \checkmark &  &  &  & \checkmark &  & \checkmark &   \\
    \cite{Paula2021} &  &  &  & \checkmark & \checkmark &  &  &  \checkmark \\
    \cite{Zhu2021} &  &  & \checkmark & \checkmark &  &  &  & \checkmark  \\
    \cite{Jalel2021} &  & \checkmark &  &  &  &  &  &  \\
    \cite{Felix2021} &  & \checkmark &  &  &  &  &  &  \\
    \rowcolor{lightgray}
    \multicolumn{9}{c}{2022}\\
    \cite{DannyGarcia2022} & \checkmark &  &  & \checkmark &  & \checkmark &  &   \\
    \cite{Guillaume2022} & \checkmark &  &  & \checkmark & \checkmark &  &  &   \\
    \cite{Lilian2022} & \checkmark &  &  & \checkmark & \checkmark &  &  &   \\
    \cite{Vincent2022} &  &  &  & \checkmark &  &  &  &   \\
    \cite{Stavropoulou2022} &  &  &  & \checkmark &  &  &  &   \\  
    \cite{Zhuang2022} & \checkmark &  &  &  &  &  &  &   \\
    \cite{AmineMasmoud2022} & \checkmark &  &  & \checkmark &  &  &  &   \\
    \cite{Ahmed2022} & \checkmark &  &  &  &  &  &  &   \\
    \cite{Payakorn2022} &  &  &  & \checkmark &  &  &  &  \\
    \rowcolor{lightgray}
    \multicolumn{9}{c}{2023}\\
    \cite{Duan2023} & \checkmark &  & \checkmark &  &  &  &  &  \\
    \cite{Yuxin2023} &  &  &  & \checkmark & \checkmark &  &  &   \\
    \cite{Peng2023} &  &  &  & \checkmark &  &  & \checkmark &   \\
    \cite{Huo2023} &  &  &  &  &  &  & \checkmark & \checkmark  \\
    \cite{Bruglieri2023} &  &  &  & \checkmark & \checkmark &  &  & \checkmark \\    
    \cite{Maryam2023} &  &  &  & \checkmark & \checkmark &  &  &   \\
    \cite{Andres2023} & \checkmark &  &  &  &  &  &  &  \\
    \cite{Qinge2023} &  & \checkmark &  &  &  &  &  &   \\
    \cite{Zakir2023} & \checkmark &  &  &  & \checkmark &  &  &   \\
    \rowcolor{lightgray}
    \multicolumn{9}{c}{2024}\\
    \cite{Torkzaban2024} &  &  &  & \checkmark &  &  &  &   \\
    \cite{Baptista2024} & \checkmark &  &  &  &  &  &  &   \\
    \cite{Vincent2024} &  &  &  & \checkmark & \checkmark &  &  &   \\
    \cite{Kim2024} & \checkmark &  &  &  &  &  &  &   \\
    \cite{Zhang2024} & \checkmark &  &  &  &  &  &  &   \\
    \cite{Rim2024} & \checkmark &  &  &  &  &  &  &   \\
    \cite{Ma2024} & \checkmark &  &  &  &  &  &  &   \\
    \cite{Pilati2024} &  &  &  & \checkmark &  &  &  &   \\
    \cite{Xiao2024} &  &  &  & \checkmark & \checkmark &  &  &  \\
    \cite{Xie2024} &  &  &  & \checkmark &  &  & \checkmark & \checkmark  \\
    \cite{Francesco2024} & \checkmark &  &  &  &  &  &  &   \\
    \cite{Abid2024} & \checkmark & \checkmark &  & \checkmark &  & \checkmark &  &   \\
    \cite{Gasset2024} & \checkmark &  &  &  &  &  &  &   \\
    \cite{Jasim2024} &  &  &  & \checkmark &  &  &  &  \\
    \cite{Zhou2024} & \checkmark &  &  & \checkmark & \checkmark &  & \checkmark &  \\ 
\botrule
\end{tabular*}
\end{table}
\section{Discussion and future directions}\label{section8}
\begin{figure}[h!]
\centering
\includegraphics[width=119mm]{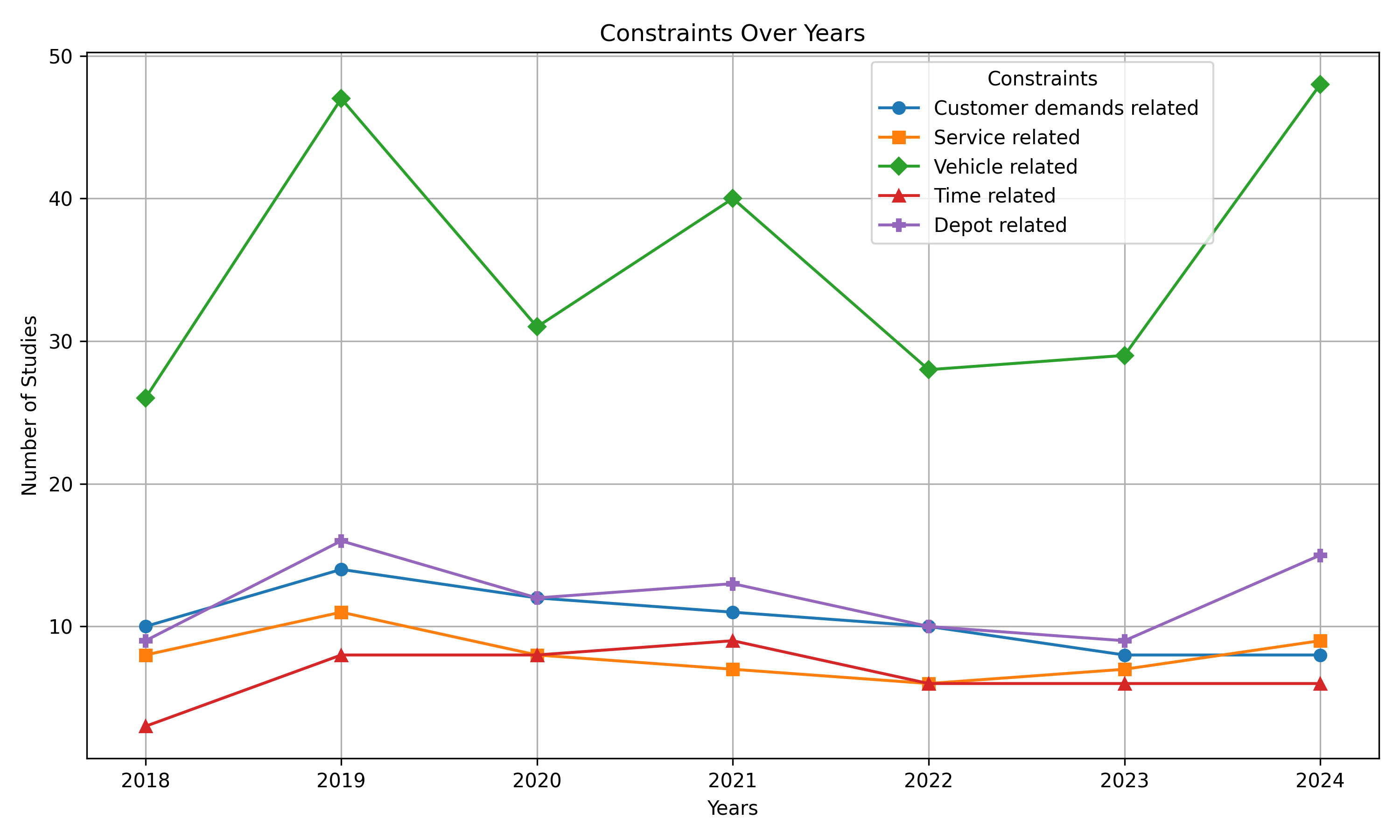}
\caption { VRP constraint use trends \label{Fig5}}
\end{figure}

Figure \ref{Fig5} depicts recent studies awareness of the five categories of constraints, namely customer demands, service, vehicle, time, and depot-related. The number of studies corresponding to each category over the period starting from $2018$ to $2024$.

\begin{figure}[h!]
\centering
\includegraphics[width=119mm]{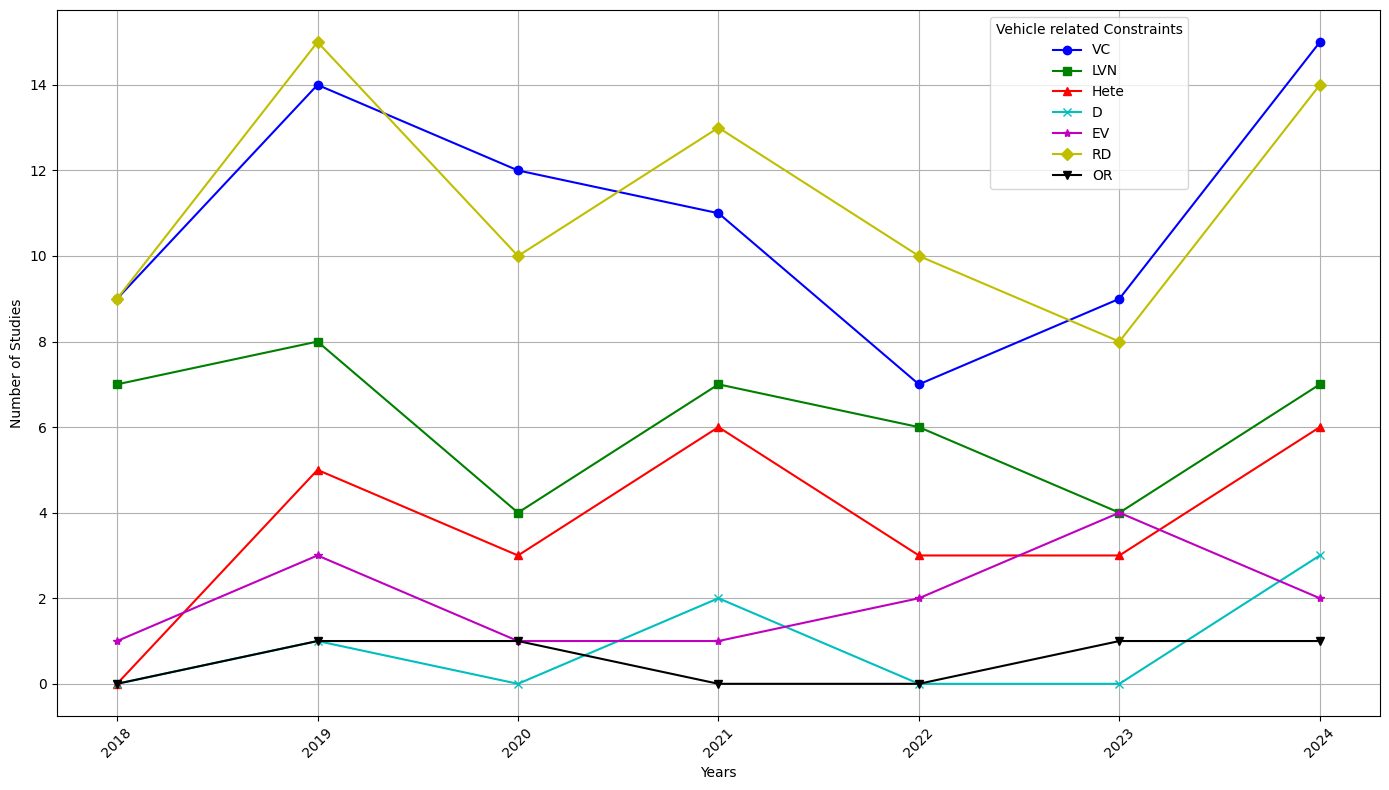}
\caption { Trends in detailed vehicle-related extensions. \label{Fig6}}
\end{figure}

Vehicle-related extensions seems to be the most extensively used category throughout the analyzed period, peaking in $2019$ with over $50$ studies. This observation underscores the significant emphasis placed by researchers on operational issues concerning vehicle logistics, which constitute a foundational aspect of VRP research. Within this category, constraints such as capacity and return to depot consistently received the highest attention, whereas less common extensions, including heterogeneity and drone utilization, exhibited periodic absences in the literature. For instance, drone utilization and open return extensions were not addressed in $2018$, and the latter was absent again in $2021$ and $2022$. These findings suggest a primary focus on traditional vehicle management challenges as shown in Figure \ref{Fig6}.

\begin{figure}[h!]
\centering
\includegraphics[width=119mm]{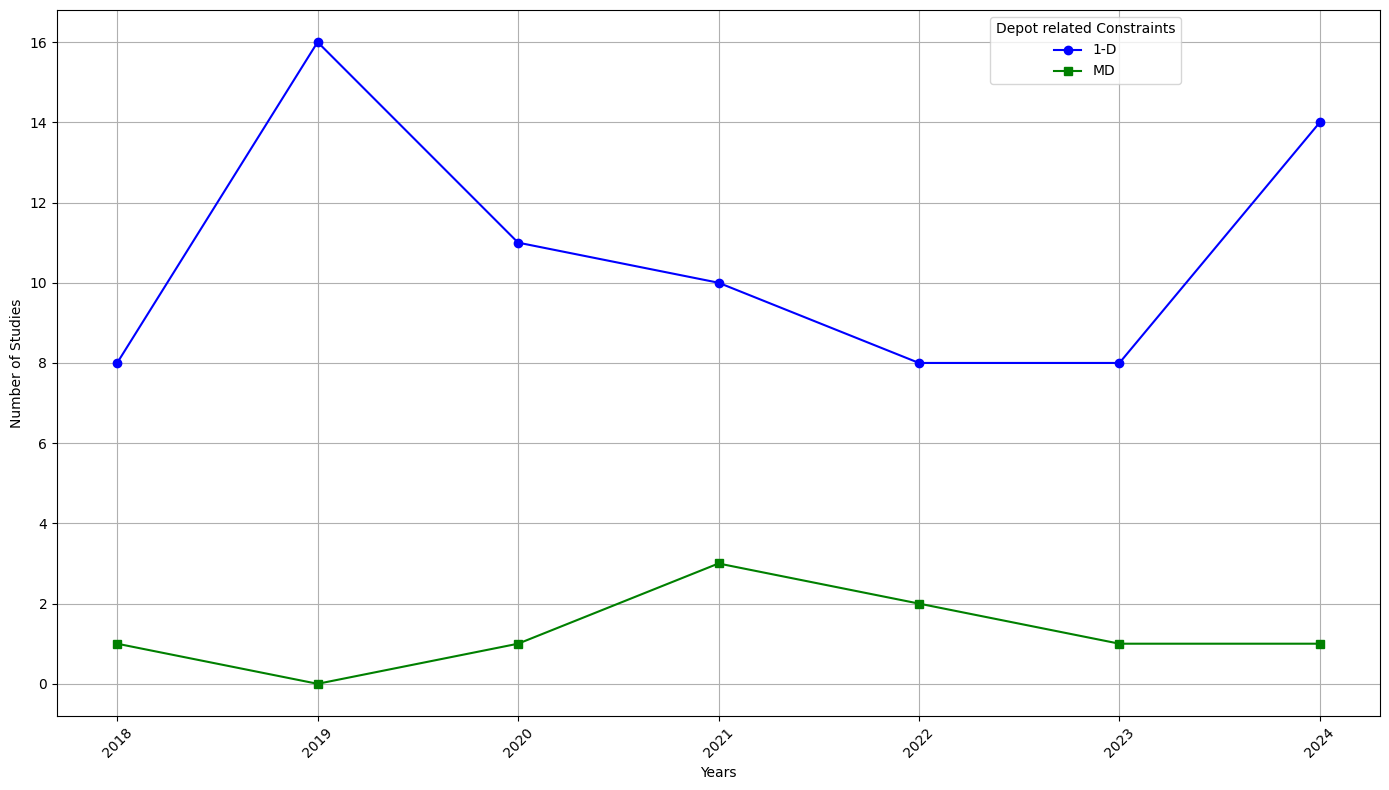}
\caption { Trends in depot related extensions \label{Fig7}}
\end{figure}

Depot-related constraints rank as the second most explored category. Figure \ref{Fig7}, reveals a research bias toward single-depot configurations. This approach has consistently dominated the literature across all years. Meanwhile, the limited number of studies addressing multiple-depot configurations highlights a significant research gap in exploring the complexities introduced by distributed depot networks.\\

\begin{figure}[h!]
\centering
\includegraphics[width=119mm]{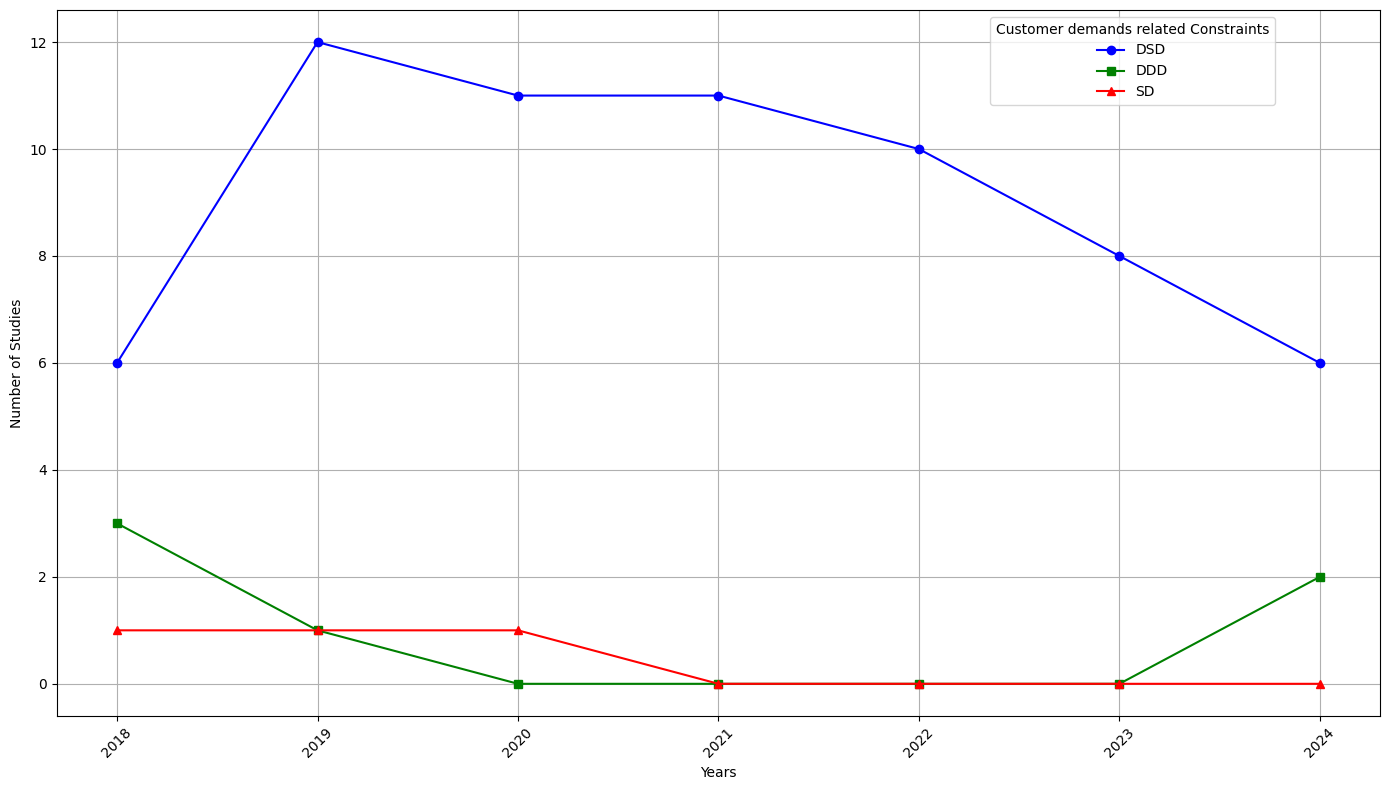}
\caption { Trends in customer demand related extensions \label{Fig8}}
\end{figure}

Customer demand related constraints rank as the Third most explored category, with a peak of studies in $2019$. This category encompasses deterministic, dynamic and stochastic demands, reflecting the diverse modeling approaches of customer requirements handling. Deterministic static demand constraints were dominant throughout all the period. In contrast, dynamic and stochastic demand constraints remained poorly explored, despite their significant relevance to practical and uncertain logistics environments, as shown in Figure \ref{Fig8}.\\

\begin{figure}[h!]
\centering
\includegraphics[width=119mm]{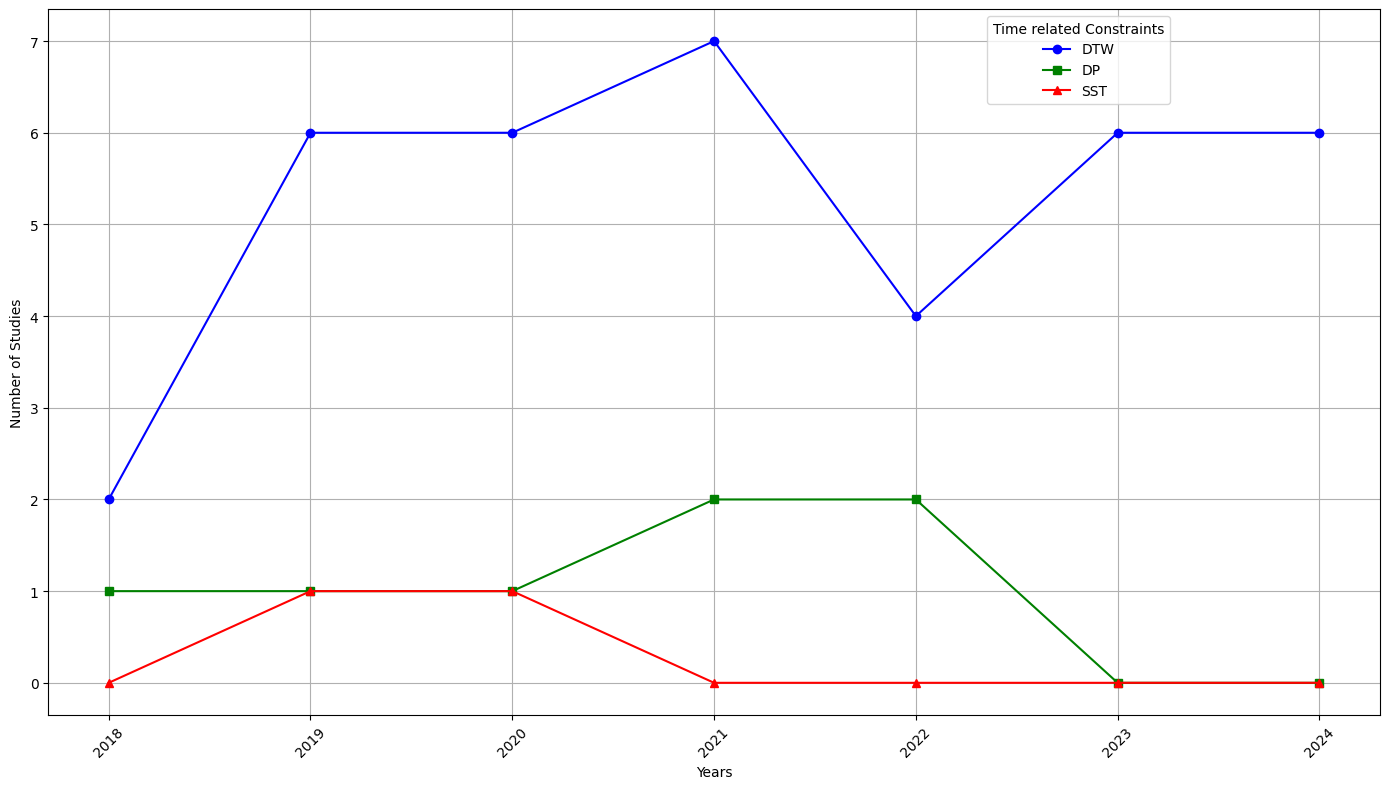}
\caption { Trends in time related extensions \label{Fig9}}
\end{figure}

Both time and service related extensions exhibit notable variations across the years. The number of studies addressing time-related constraints increased significantly between $2019$ and $2021$, with deterministic time window constraints consistently emerging as the most dominant type as illustrated in Figure \ref{Fig9}. This highlights the critical importance of aligning delivery schedules with customer availability.\\

\begin{figure}[h!]
\centering
\includegraphics[width=119mm]{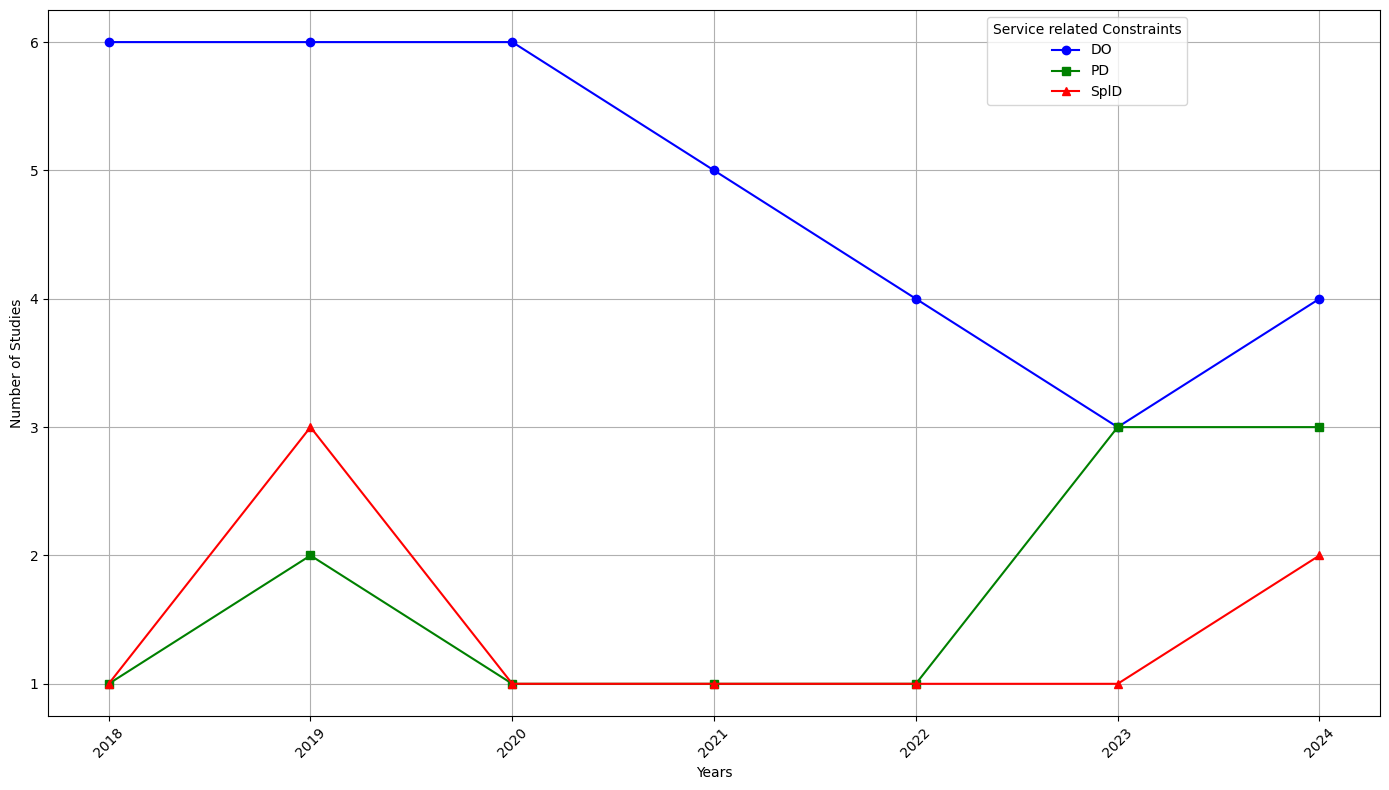}
\caption { Trends in service related extensions \label{Fig10}}
\end{figure}

In contrast, service-related extensions were most prominent in $2018$ and $2024$. Delivery-only service assumption emerged as the most utilized type throughout the study period. However, it exhibited a clear decline from $2020$ on. Conversely, pickup-and-delivery and split-delivery extensions were absent in many years, including $2018, 2020, 2021$, and $2022$ for both types, and $2023$ for split-delivery, as illustrated in Figure \ref{Fig10}.\\

Besides, analyzing VRP variants highlights several key trends and research gaps across different categories of constraints and configurations, as we noticed through Figures \ref{Fig14} to \ref{Fig19}.

\begin{figure}[h!]
\centering
\includegraphics[width=119mm]{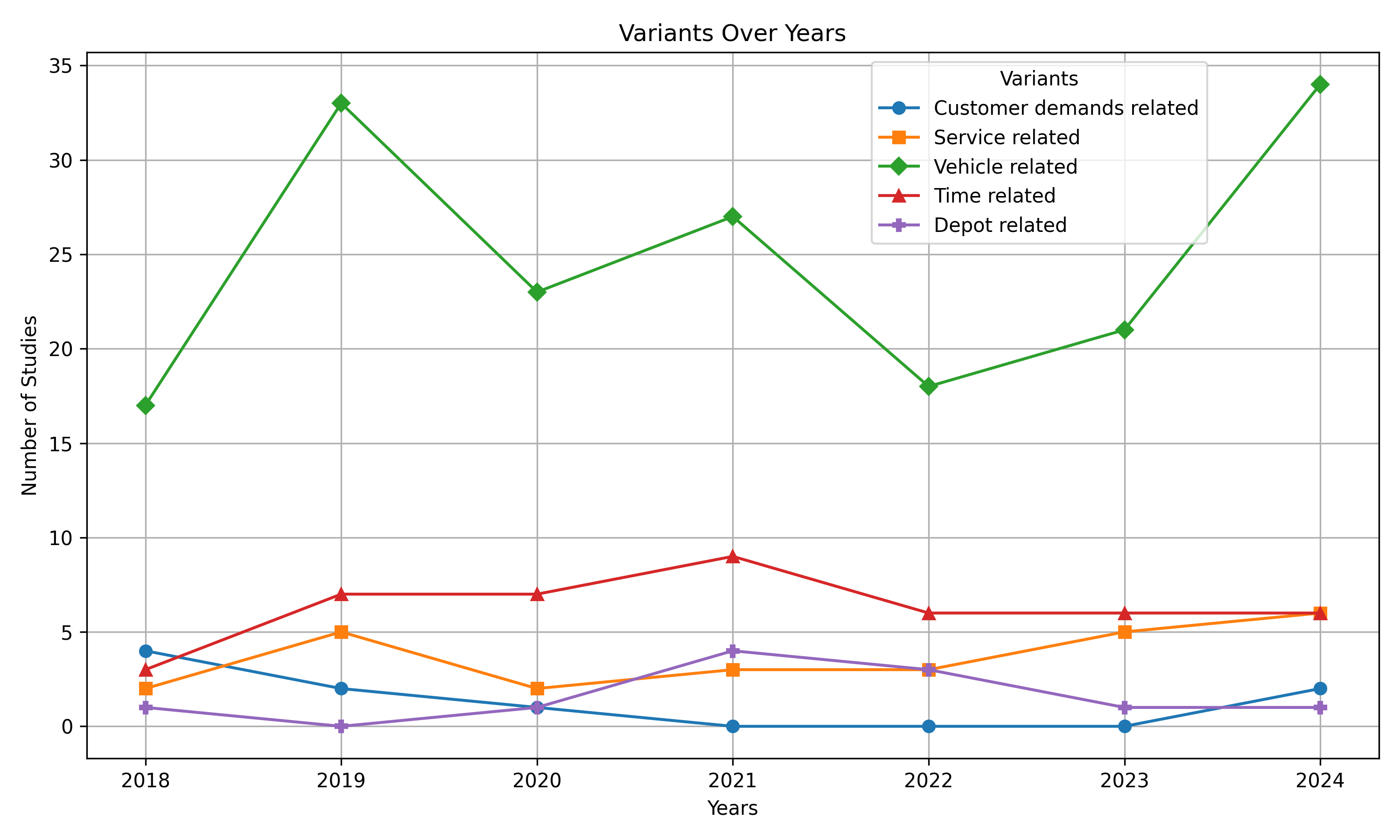}
\caption { Trends in VRP variants \label{Fig11}}
\end{figure}

\begin{figure}[h!]
   \centering
   \includegraphics[width=119mm]{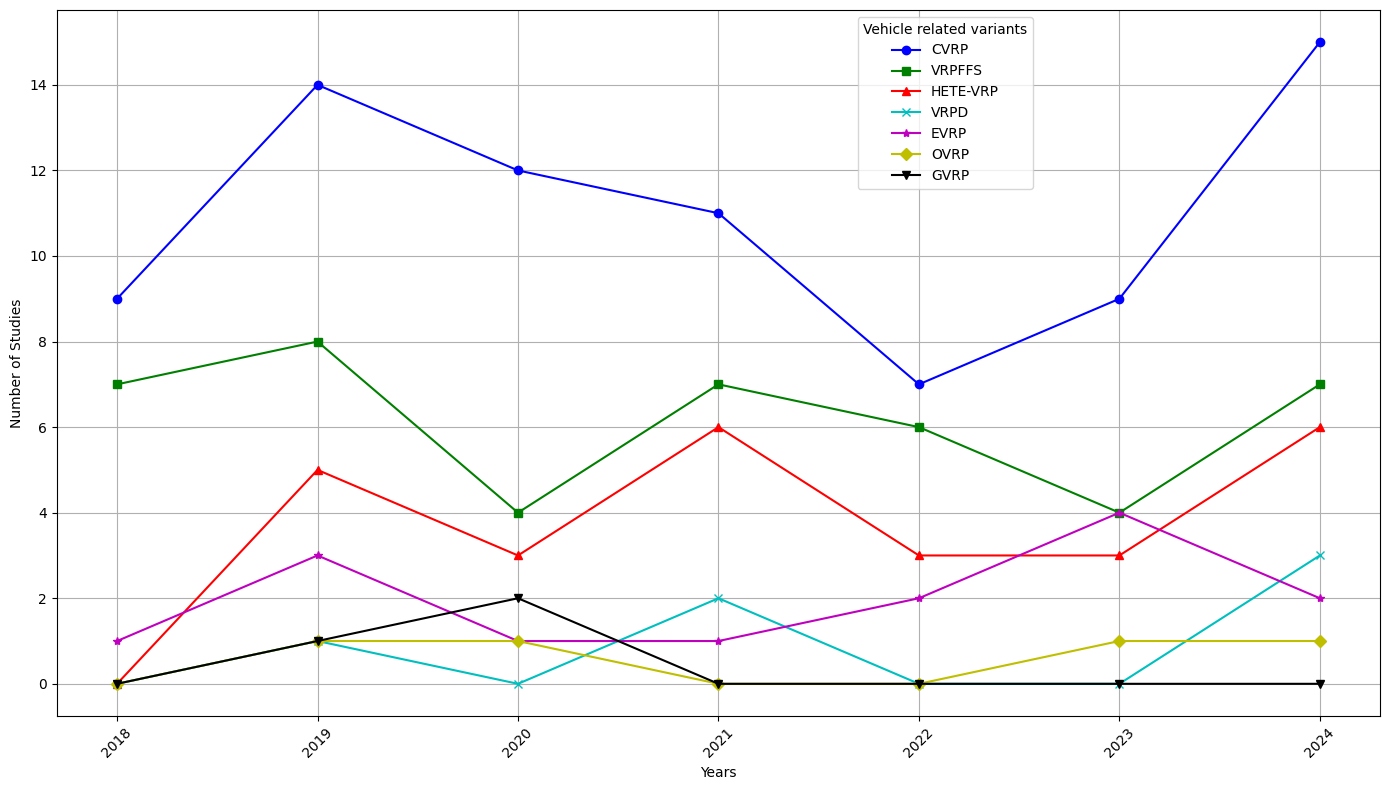}
   \caption { Trends in vehicle related variants \label{Fig12}}
\end{figure}

Vehicle related variants constitute the most extensively researched category throughout the study period. Within this domain, the Capacitated Vehicle Routing Problem stands out as the most dominant variant, consistently attracting significant attention from researchers. Other vehicle-related variants, such as the Vehicle Routing Problem with Fixed Fleet Size exhibit variable interest levels across the years. However, emerging and environmentally significant variants like the Electric VRP, Green VRP, Open VRP, and VRP with Drones remain significantly under-explored. This trend underscores a research gap in addressing modern challenges in vehicle routing, as illustrated in Figure \ref{Fig12}.\\

\begin{figure}[h!]
   \centering
   \includegraphics[width=119mm]{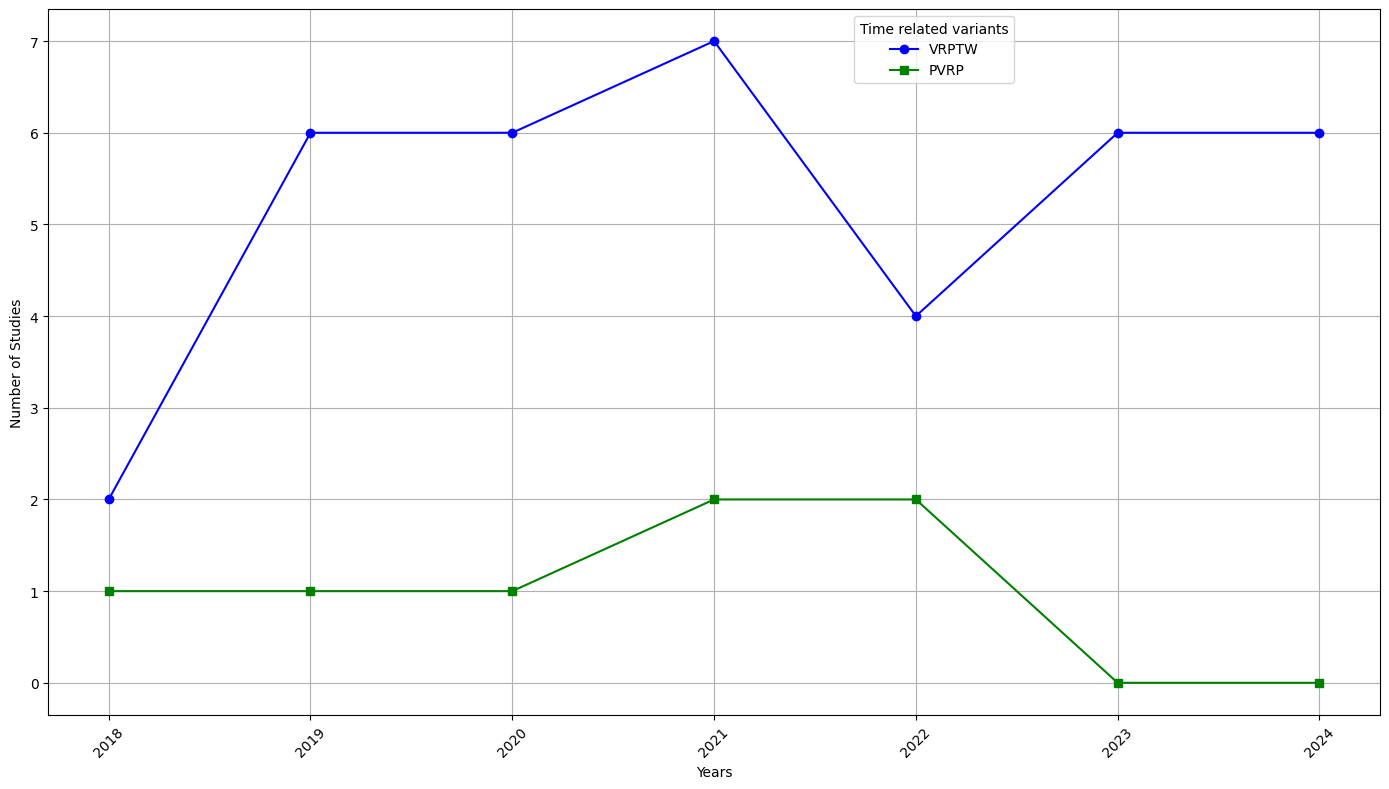}
   \caption{ Trends in time related variants \label{Fig13}}
\end{figure}

\begin{figure}[h!]
   \centering
   \includegraphics[width=119mm]{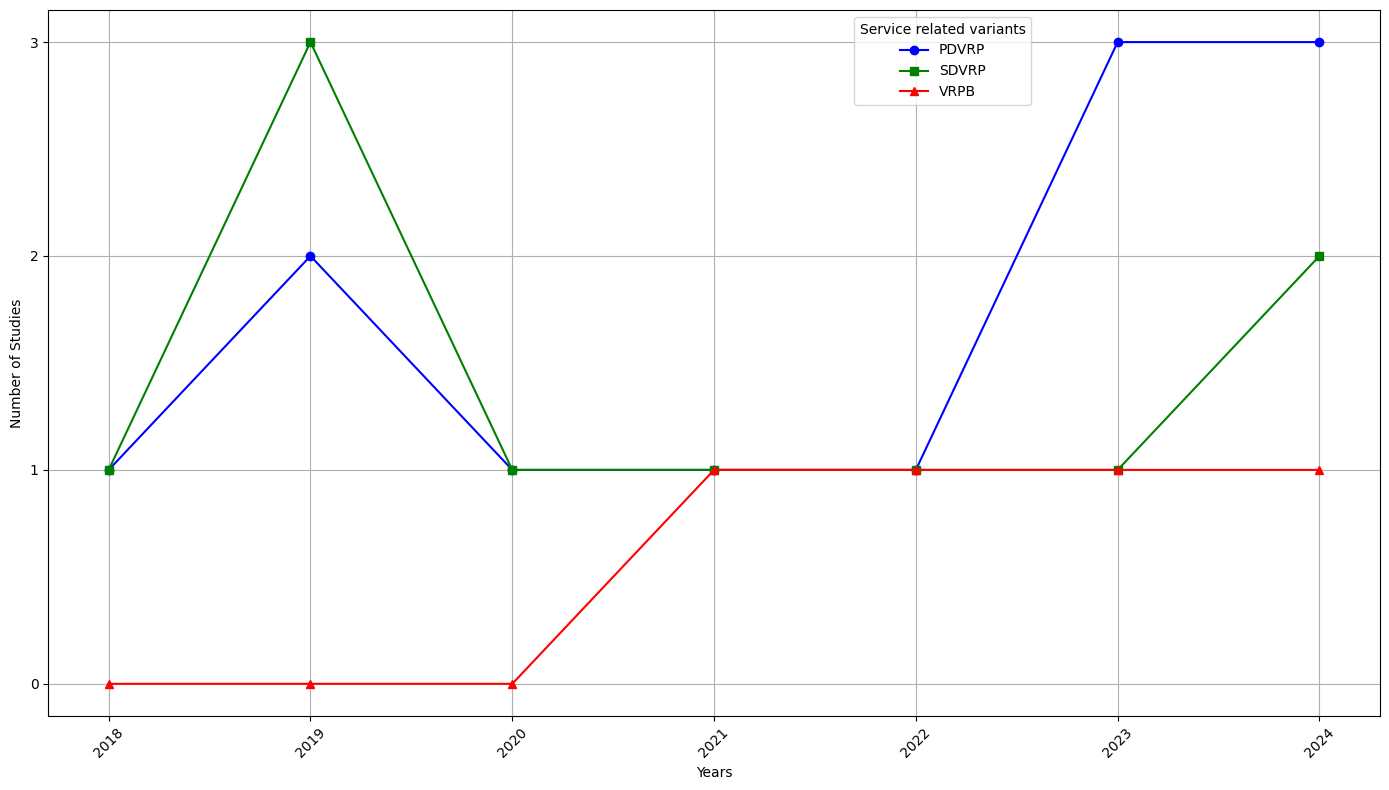}
   \caption { Trends in service related variants \label{Fig14}}
\end{figure}

\begin{figure}[h!]
   \centering
   \includegraphics[width=119mm]{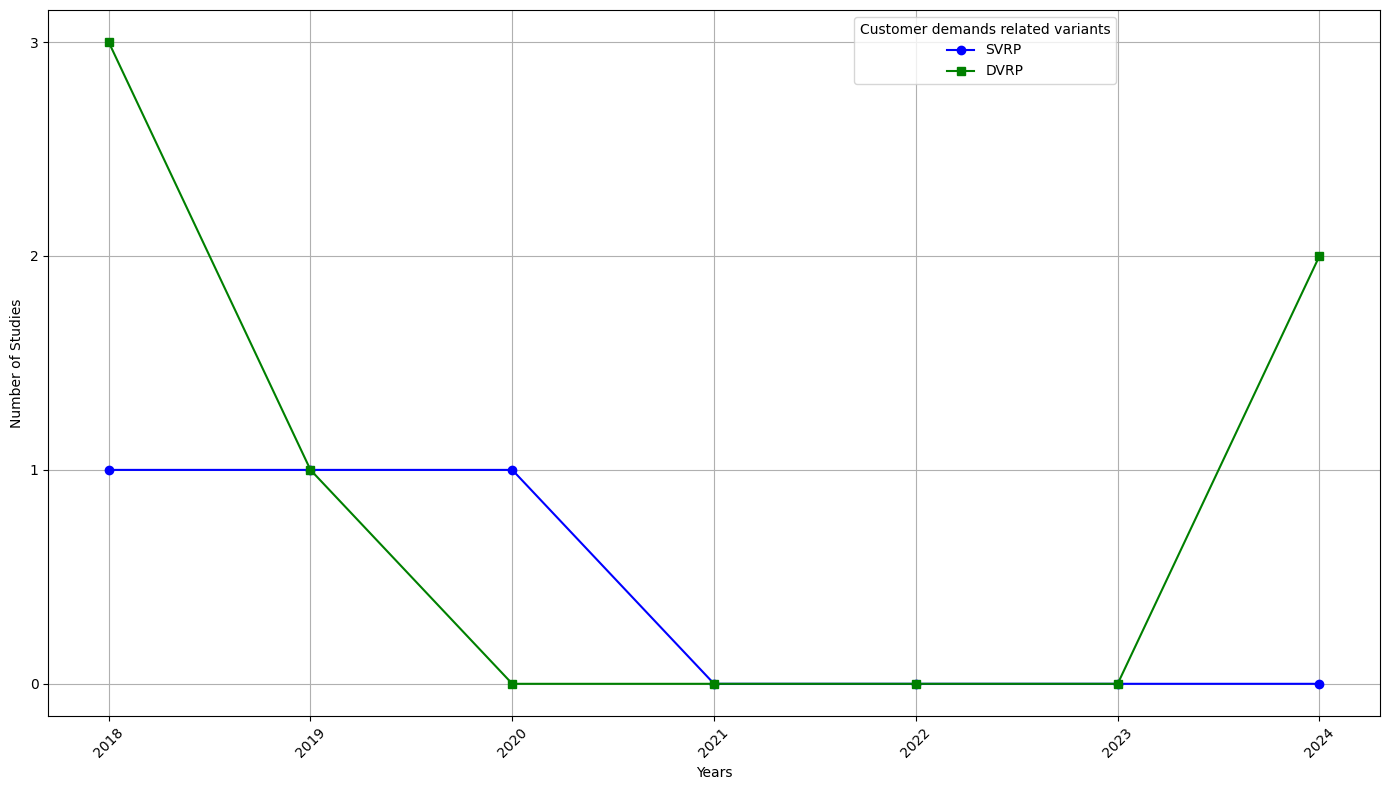}
   \caption { Trends in customer demands related variants \label{Fig15}}
\end{figure}

\begin{figure}[h!]
   \centering
   \includegraphics[width=119mm]{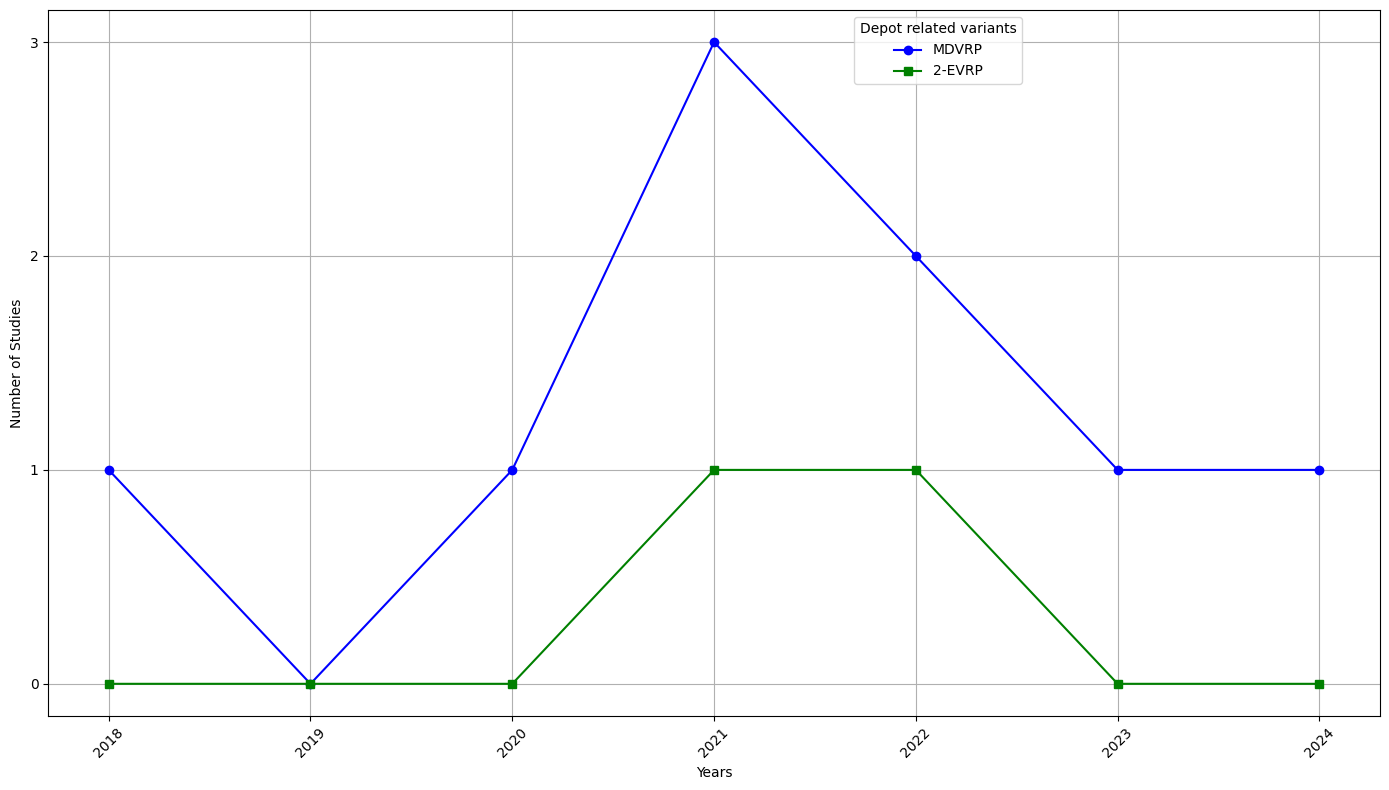}
   \caption{ Trends in depot related variants \label{Fig16}}
\end{figure}

\begin{figure}[h!]
   \centering
   \includegraphics[width=119mm]{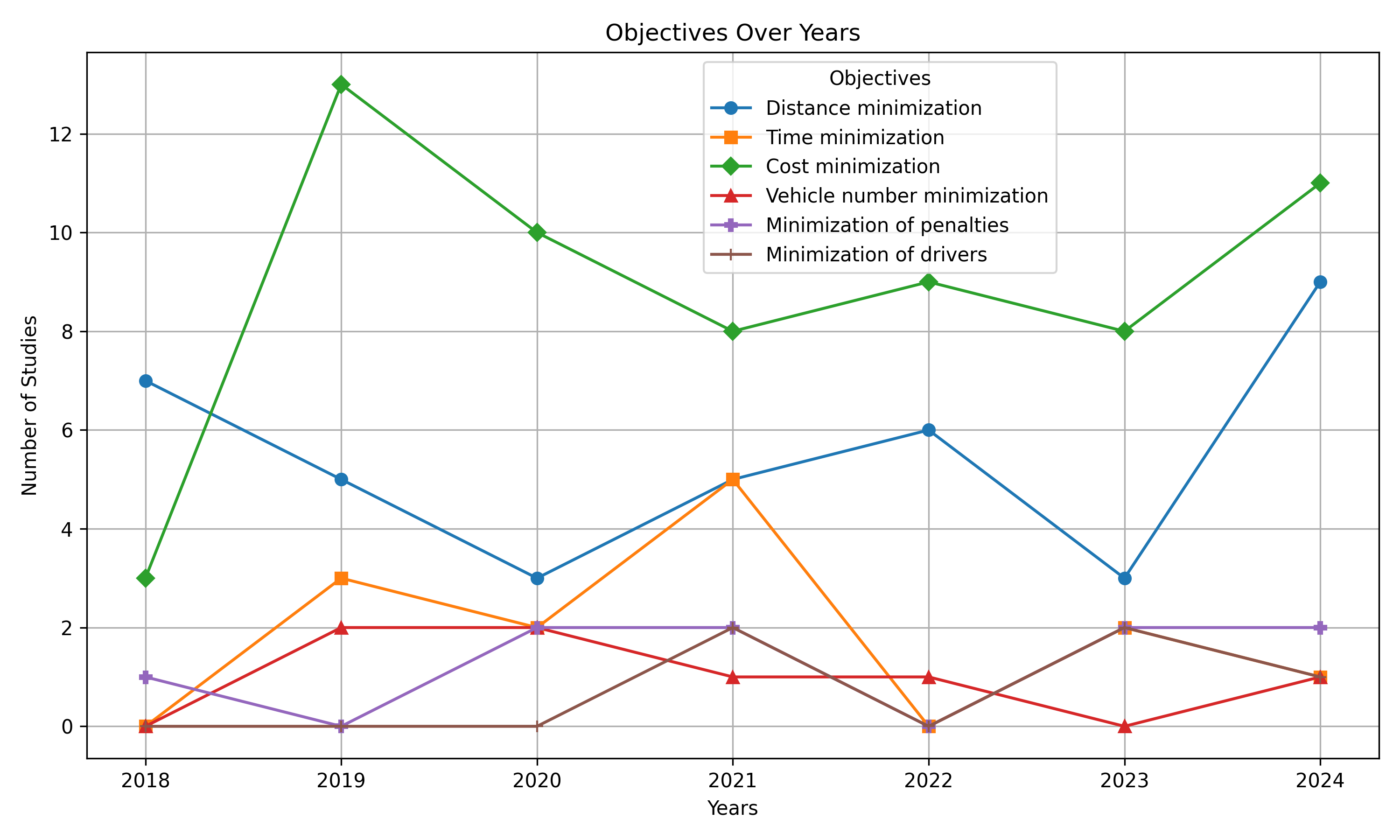}
   \caption{ Trends in VRP objectives \label{Fig17}}
\end{figure}

\begin{figure}[h!]
   \centering
   \includegraphics[width=119mm]{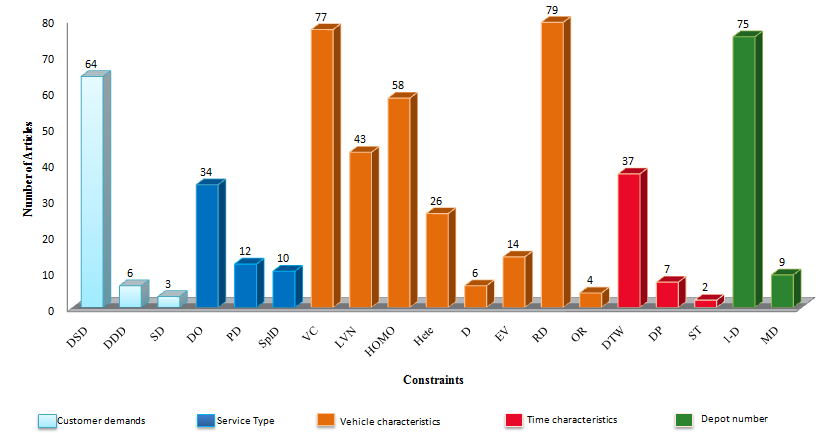} 
   \caption{Distribution of VRP research works according to constraints \label{Fig18}}
\end{figure}

\begin{figure}[h!]
   \centering
   \includegraphics[width=119mm]{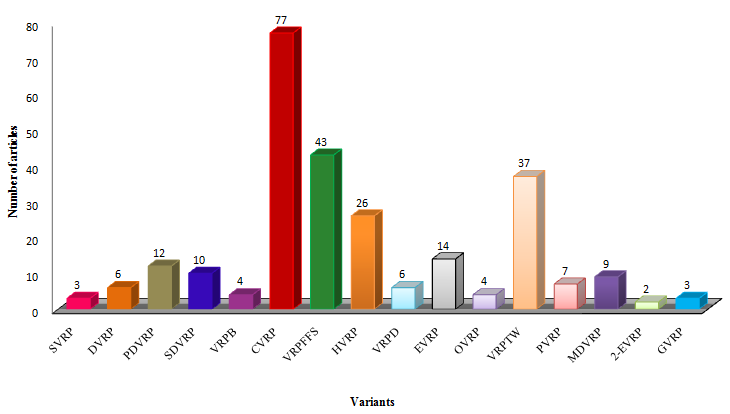}
   \caption{Distribution of VRP research works variants \label{Fig19}}
\end{figure}

Time related variants rank as the second-most studied category. Within this group, the Time Windows VRP is the predominant focus, reflecting its practical importance in aligning delivery schedules with customer availability. Conversely, the Periodic Vehicle Routing Problem has garnered less attention, as shown in Figure \ref{Fig13}.\\

We notice a varying dominance among the remaining three categories. Research in service related variants exhibits fluctuating trends. Figure \ref{Fig14} reveals that the two most extensively studied variants in this category are the Stochastic Vehicle Routing Problem and the pickup-and-delivery Vehicle Routing Problem.

Customer demands related variants category reflects also a varying levels of interest over the years. For instance, the Dynamic VRP was the most researched variant in both $2018$ and $2024$. On the other hand, the Stochastic VRP was prominent in 2020 but completely absent from studies in $2021$, $2022$, and $2023$. This discontinuity in research suggests a potential oversight in addressing uncertain or probabilistic demand scenarios, which are critical in practical logistics applications, as shown in Figure \ref{Fig15}.

Research in depot-related variants is relatively scarce. The Multi-Depot VRP dominates this segment, highlighting the emphasis on managing multiple depots for optimized routing. However, other configurations, such as the Two-Echelon VRP, are rarely studied, with only single occurrences reported in $2021$ and $2022$, as shown in Figure \ref{Fig16}.\\

In another hand, the analysis of the trends for the different VRP objectives used in our sample, as shown in Figure \ref{Fig17}, indicates that the most frequently used objective is cost minimization, which has been predominant in most years. However, there is a noticeable increase in emphasis on this objective over time, with a slight decline observed in $2020$, $2021$ and $2023$. Following cost optimization, the distance minimization objective has also been significant throughout the years.

\begin{figure}[h!]
   \centering
   \includegraphics[width=119mm]{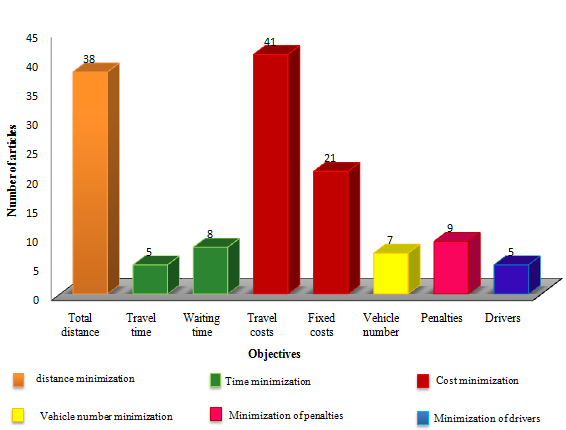}
   \caption{ Distribution of VRP research objectives \label{Fig20}}
\end{figure}

In contrast, the objective of vehicle number minimization shows a lack of research works, with no studies addressing this objective in both $2018$ and $2023$. Additionally, minimization of penalties and drivers are present in very few papers published during the study years. This highlights critical gaps in the research landscape regarding optimizing vehicle amount, penalties and number of drivers objectives within recent VRP studies. \\
Recapitulating what has been noticed so far, we can say that according to the analysis of recent research explored in this paper, the most frequently studied constraints in the VRP literature involve deterministic static customer demands, along with fixed travel and service times. A single depot setup, combined with delivery-only services, is also common. Additionally, many studies incorporate a limited fleet of vehicles with fixed capacities, where the vehicles must begin and conclude their routes at the depot, often with time windows as an additional constraint, as illustrated in Figure \ref{Fig18}.
Consequently, the most commonly studied VRP variants include the CVRP, VRPFFS, VRPTW and HVRP, as depicted in Figure \ref{Fig19}. \\
% \\
For the objectives, our study shows that the most frequently used criteria are either minimizing the total distance travelled or reducing costs (see Figure \ref{Fig20}). These cost optimizations can take various forms, depending on the specific problem context.\\

As future directions, we think that incorporating dynamic and stochastic characteristics into customer demands and service times is essential to better reflect the uncertainties inherent in real-world applications. Additionally, utilizing multiple depots is crucial for enhancing delivery operations, particularly for large companies with extensive branch networks. Accommodating both delivery and pickup operations, along with managing split deliveries, is also a promising approach for future research. Moreover, integrating  heterogeneous fleet of vehicles significantly enhances the applicability of solutions to real-world scenarios, allowing for greater flexibility in routing without the strict requirement for vehicles to return to the depot after each route. Furthermore, the adoption of recent technologies, such as electric vehicles and drones, presents promising avenues for development, to cope with international recommendations such as using green logistics aiming at preserving the environment and reducing pollution.

Besides, optimizing multiple objectives will be a valuable approach for future research, enabling decision-makers to balance competing factors, such as minimizing costs while maximizing service quality and customer satisfaction.\\

Finally, this study is based on the analysis of $84$ articles published between $2018$ and $2024$, which explore various VRP constraints, variants, and objectives . While the sample provides valuable insights, it is limited in scope and may not fully represent the diversity of VRP research across different domains. Moreover, the study primarily focuses on some common constraints, variants, and objectives presented in the selected sample. Expanding the sample to include a broader range of articles could potentially provide a more comprehensive understanding of the field.

\section{Conclusion}\label{section9}
This article thoroughly explores the Vehicle Routing Problem (VRP) by discussing its various constraints, objectives and variants present in recent research.

Our goal is to enhance VRP understanding and provide a solid foundation for researchers in this field. By offering a comprehensive overview, we aim to support further research and development, deepening the mastering of VRP's complexities.

The choice of a specific variant and its associated constraints depends on the detailed nature of the problem being addressed. Additionally, the selection of the objective must extend beyond simple travel distance optimization. Critical factors such as reducing waiting times, optimizing energy consumption while insuring customer satisfaction and nature-friendly approaches must have a decisive role on the solution to be adopted.

Furthermore, the complexity of the VRP increases with the introduction of additional parameters. Indeed, considering dynamic changes such as stochastic demands further complicates the problem, necessitating more sophisticated approaches to refine the model and its solving approach.

Incorporating these diverse parameters along with many objectives and constraints to construct models that are tailored to cope with real-world challenges leading to more effective solutions is the most needed work toward which further research efforts must be directed.

\bibliography{sn-bibliography}% common bib file

\end{document}